%% file: aipsamp.tex
\documentclass[%
 aip,
 amsmath,amssymb,
 reprint,%
]{revtex4-1}

\usepackage{graphicx}% Include figure files
\usepackage{dcolumn}% Align table columns on decimal point
\usepackage{bm}% bold math
\usepackage[utf8]{inputenc}
\usepackage[T1]{fontenc}
\usepackage{mathptmx}
\usepackage{etoolbox}

\usepackage{pbox}
\usepackage{physics}
\usepackage{bm}
\usepackage{array}
\usepackage{booktabs}
\usepackage{multirow}
\usepackage{cellspace}
\usepackage{verbatim}
\usepackage{dsfont}

\usepackage{placeins}

\usepackage{natbib}
\usepackage[caption=false, position=top]{subfig}
\usepackage{xcolor}
\usepackage{siunitx}

\newcommand{\srad}{A_\mathrm{rad}}
\newcommand{\sdip}{A_\mathrm{dip}}

\newcommand{\CNOT}{\textsc{cnot}}

\newcommand{\NOT}{\textsc{not}}

\newcommand{\CPMGXY}{\textsc{cpmg}\textsubscript{\textsc{xy}}}
\newcommand{\UR}{\textsc{ur}}
\newcommand{\UTOF}{U_\textsc{tof}}

\makeatletter
\def\@email#1#2{%
 \endgroup
 \patchcmd{\titleblock@produce}
  {\frontmatter@RRAPformat}
  {\frontmatter@RRAPformat{\produce@RRAP{*#1\href{mailto:#2}{#2}}}\frontmatter@RRAPformat}
  {}{}
}%
\makeatother
\begin{document}

\preprint{AIP/123-QED}

% \title[Sample title]{Sample Title:\\with Forced Linebreak}
% % Force line breaks with \\
% \author{A. Author}
%  \altaffiliation[Also at ]{Physics Department, XYZ University.}%Lines break automatically or can be forced with \\
% \author{B. Author}%
%  \email{Second.Author@institution.edu.}
% \affiliation{ 
% Authors' institution and/or address%\\This line break forced with \textbackslash\textbackslash
% }%

% \author{C. Author}
%  \homepage{http://www.Second.institution.edu/~Charlie.Author.}
% \affiliation{%
% Second institution and/or address%\\This line break forced% with \\
% }%

\title{
    Classical Half-Adder using Trapped-ion Quantum Bits:\\
    Towards Energy-efficient Computation
}

\author{Sagar Silva Pratapsi}
\email{spratapsi@tecnico.ulisboa.pt}
\altaffiliation{Corresponding author.}
\affiliation{
Instituto Superior Técnico, University of Lisbon,
Lisbon 1049-001, Portugal.
}
\affiliation{Instituto de Telecomunicações,
Lisbon 1049-001, Portugal.}

\author{Patrick H. Huber}
\email{p.huber@physik.uni-siegen.de}
\altaffiliation{Corresponding author.}
\affiliation{Department of Physics, School of Science and Technology, University of Siegen, 57068 Siegen, Germany}

\author{Patrick Barthel}
\affiliation{Department of Physics, School of Science and Technology, University of Siegen, 57068 Siegen, Germany}

\author{Sougato Bose}
\affiliation{Department of Physics and Astronomy, University College London, London WC1E 6BT, UK}

\author{Christof Wunderlich}
\email{Christof.Wunderlich@uni-siegen.de}
\altaffiliation{Corresponding author.}
\affiliation{Department of Physics, School of Science and Technology, University of Siegen, 57068 Siegen, Germany}

\author{Yasser Omar}
\email{contact.yasser@pqi.pt}
\altaffiliation{Corresponding author.}
\affiliation{
Instituto Superior Técnico, University of Lisbon,
Lisbon 1049-001, Portugal.
}
\affiliation{Physics of Information and Quantum Technologies group, Center of Physics and Engineering of Advanced Materials (CeFEMA),
Lisbon 1049-001, Portugal}
\affiliation{PQI -- Portuguese Quantum Institute,
Lisbon 1600-531, Portugal}

\date{\today}% It is always \today, today,
             %  but any date may be explicitly specified

\begin{abstract}
Reversible computation has been proposed as a future paradigm for energy efficient computation, but so far few implementations have been realised in practice.
Quantum circuits, running on quantum computers, are one construct known to be reversible. In this work, we provide a proof-of-principle of classical logical gates running on quantum technologies.
In particular, we propose, and realise experimentally, Toffoli and Half-Adder circuits suitable for classical computation, using radiofrequency-controlled $^{171}$Yb$^+$ ions in a macroscopic linear Paul-trap as qubits.
We analyse the energy required to operate the logic gates, both theoretically and experimentally,
with a focus on the control energy.
We identify bottlenecks and possible improvements in future platforms for energetically-efficient computation,
\textit{e.g.}, trap chips with integrated antennas and cavity \textsc{qed}.
Our experimentally verified energetic model also fills a gap in the literature of the energetics of quantum information, and outlines the path for its detailed study, as well as its potential applications to classical computing.
\end{abstract}

\maketitle

%%%%%%%%%%%%%%
% Introduction
%%%%%%%%%%%%%%

Computational tasks are responsible for a non-negligible part of the world's energy consumption.
It is estimated that computationally-intensive data-centres represent 1\% of the global energy budget \cite{masanet2020recalibrating}.
So far, increases in energy efficiency have been able to offset the growing demand for computation:
peak-usage energy efficiency has doubled every 1.5 years during the 1960--2000 period, while since the 2000s this figure is closer to 2.6 years \cite{masanet2020recalibrating, naffzigerenergy}.
However, processor efficiency gains cannot continue to grow forever. There is a fundamental limitation of the current paradigm of non-reversible computation, known as Landauer's principle \cite{landauer1961irreversibility}, where each irreversible bit operation dissipates $k_B T\ln 2$ of heat.

Reversible computation may thus become an important computation paradigm in the future. Reversible systems may also avoid the heat costs of contemporary \textsc{cmos} processors, such as capacitor charging, switching and current leakage \cite{rabaey_low_1996, koomey2015primer}, which are ultimately responsible for the typical 40\% energy cost for cooling in data centres \cite{ni_review_2017}; they may also protect against external attacks such as power usage analysis.
It is, then, worthwhile to investigate how energy-efficient reversible platforms can become.
Some proposals for reversible computing platforms have been
billiard-ball models \cite{fredkin2002design, fredkin1982conservative},
adiabatic circuits \cite{seitz1985hot, koller1992adiabatic, hall1992electroid, merkle_practical_reversible_logic, merkle1993reversible},
nano-machines \cite{drexler1991molecular, drexler1992nanosystems, merkle2016molecular, merkle2018mechanical, hogg2017evaluating},
superconducting devices \cite{likharev1977dynamics, hosoya1991quantum, ren2011progress},
quantum-dot cellular automata \cite{lent1994quantum},
and others (see \cite{frank2018review} for a review of reversible computation). But, so far, experimental realisations of reversible computation are lacking in practice.
Quantum mechanical systems, which evolve unitarily, are also reversible by nature, and are thus an attractive candidate for energetically efficient computation \cite{benioff1980computer, benioff1982}.
Although quantum platforms are limited by coherence time, we can reset the coherence for classical computations by measuring in the computational basis in-between logical operations. We may also exploit super-selection rules to protect classical information, as was proposed recently in a quantum dot platform \cite{moutinho2022quantum}.
Can we then build energy efficient circuits for universal reversible computation using quantum computing platforms?

\begin{figure*}[t]
    \includegraphics{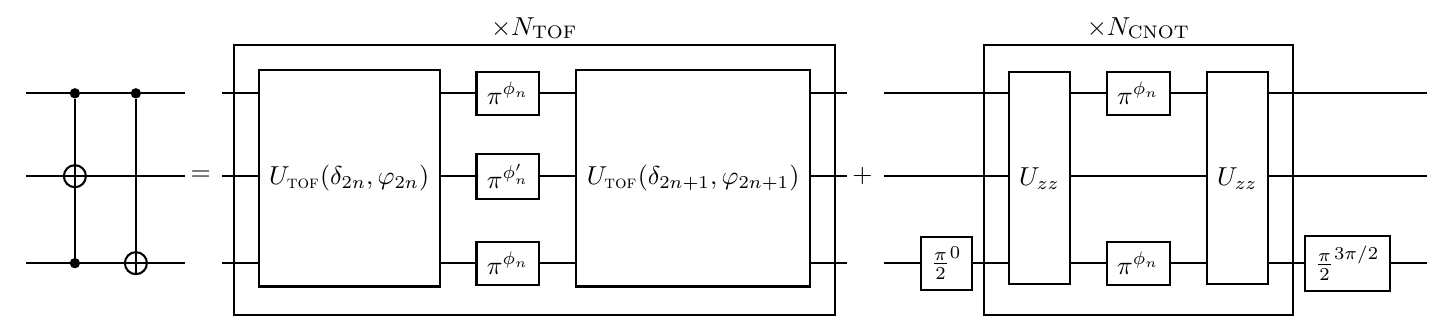}
    \caption{
    \textbf{A Half-adder circuit} using a Toffoli followed by a \CNOT{} gate. We choose the central qubit as the target of the Toffoli gate to fullfil the condition $J_{12}=J_{23}$ from Equation~\eqref{eq:HTOF}. 
    \textbf{The Toffoli gate} decomposes into a unitary $\UTOF{}(\delta_n,\phi_n)$ (generated from Hamiltonian~\eqref{eq:Hrotating} for $\SI{14.9}{\ms}/400$) and single qubit $\pi$-pulses with some phase $\phi$ ($\pi^\phi$, implementing Dynamical Decoupling). The block is repeated $N_\text{TOF}=200$ times with updated values of $\delta_n$, $\varphi_n$ and DD phases $\phi_n$ and $\phi'_n$. The latter are chosen to implement a universal robust DD sequence on qubits 1 and 3 and a \CPMGXY{} on qubit 2. The values $(\delta_{2n}, \delta_{2n+1})$ alternate between $(\delta, -\delta)$ and $(-\delta, \delta)$ for each $\pi$-pulse, while $(\varphi_{2n}, \varphi_{2n+1})$ alternates between $(0, \pi)$ and $(\pi, 0)$ for each $\pi^{\pi/2}$-pulse.
    \textbf{The CNOT gate} decomposes into a $U_{zz}$ gate (implementing the $zz$ coupling) and single qubit $\pi$-pulses. The block is repeated $N_\text{CNOT}=120$ times. The phases $\phi_n$ implement a UR DD sequence on the control and target qubits.}
    \label{fig:half-adder}
\end{figure*}

In this work, we explore an implementation of reversible computation using quantum technologies, by realising a classical Half-Adder circuit---an important building block for arithmetic operations\cite{mano1997logic}---using quantum states of trapped ions.
To do so, we implement a Toffoli gate, itself a universal gate for classical computation.
We determine the energy to operate these gates, both theoretically and experimentally, with a special focus on the energy required to activate and control the logical gates, focusing on the power delivered to the Quantum Processing Unit (QPU), as defined later.
We point out possible improvements towards energy efficient computation.
Some works \cite{jaschke2022quantum} require realistic estimates for the energy consumed by quantum computers.
Thus, our energetic analysis, supported by experimental measurements, also fills a gap in the literature, and establishes a baseline for additional research towards understanding the energetic impact of quantum technologies \cite{PRXQuantum.3.020101}.

%%%%%%%%%%%%%%
% Gate proposals
%%%%%%%%%%%%%%
A Half-Adder circuit is a fundamental component of arithmetic circuits. It computes the logical \textsc{and} (multiplication modulo 2) and \textsc{xor} (addition modulo 2) of two input bits. It is a building block for the Full-Adder circuit, addition circuits in their ripple-carry and carry-lookahead variants, multiplier circuits and other tasks in contemporary computer processors.
The core operation behind our Half-Adder circuit is a quantum Toffoli gate, followed by the application of a \CNOT{} to the two control qubits of the Toffoli (FIG.~\ref{fig:half-adder}).
A Toffoli gate, or a controlled-controlled-\NOT{} gate, is a universal three-bit operation, \textit{i.e.}, it is sufficient to construct any classical reversible circuit.
Antonio \textit{et al.} proposed a Toffoli gate suitable for classical computation \cite{antonio_classical_2015}, which can be realised on any three-qubit physical system with constant nearest-neighbour Ising couplings, via the Hamiltonian 
\begin{equation}
    \label{eq:HTOF}
    H_\text{TOF} =
        \frac{\hbar J}{2} \left(\sigma^z_1 \sigma^z_2 + \sigma^z_2 \sigma^z_3 \right)
        + \frac{\hbar \delta }{2}\sigma^z_2
        + \frac{\hbar \Omega}{2} \sigma^x_2.
\end{equation}
Here, $\sigma_j^i$ is the $\sigma^i$ Pauli operator acting on the $j$-th qubit, appropriately tensored with the identity operators on the other qubits. The real constants $J$, $\delta$ and $\Omega$ define interaction strengths.
We simulated numerically the time evolution under the Hamiltonian~\eqref{eq:HTOF} for a time of $\pi / \Omega$ and $\delta = 2J$. We found that $\Omega \approx 1.1J$ allows for a $\approx 99\%$ classical Toffoli gate fidelity while minimising the gate time (see Supplemental Material).

%%%%%%%%%%%%%%
% Experimental implementation
%%%%%%%%%%%%%%
Ions confined in a linear Paul trap are natural candidates to implement the Hamiltonian~\eqref{eq:HTOF} \cite{Wunderlich2002,  antonio_classical_2015}.
We use  $^{171} \text{Yb}^+$ ions confined in a linear Paul trap, with a superimposed static magnetic field gradient \cite{khromova2012designer}.
The qubit states $\ket 0$ and $\ket 1$ are the two hyperfine states of the electronic ground state $^2S_{1/2}$ with total angular momentum quantum number and magnetic quantum number $\ket{F, m_F} = \ket{0,0}$ and $\ket{1, 1}$, connected by a magnetic dipole resonance near $2\pi \times \SI{12.6}{\GHz}$.
The $\ket 1$ state is sensitive to the magnetic field, which is position dependent, shifting individually the ions' resonances and, thus, allowing for individual addressing by tuning the microwave field driving the qubit resonance \cite{Piltz2014}
For high fidelity single qubit rotations, the ion crystal is cooled close to its motional groundstate using a sympathetic side-band cooling\cite{SBC_MO}.

When irradiating the ions with a microwave field with phase $\phi$ and frequency $\omega_x$, nearly resonant with the frequency $\omega_2$ of qubit 2, the ionic qubits are subject to the Hamiltonian
\begin{equation}
    H^{(i)} =
        \underbrace{\sum_{i \neq j}\frac{\hbar J_{ij}}{2} \sigma^z_i \sigma^z_j}_{H_{zz}}
        + \sum_j \frac{\hbar \omega_j \sigma^z_j}{2}
        + \hbar \Omega \cos(\omega_x t + \phi) \sigma^x_2.
    \label{realistic_hamiltonian}
\end{equation}
Here, $\omega_i$ is the resonance frequency of the $i$\textsuperscript{th} ion.
The two-qubit couplings $J_{ij}$ in a magnetic field gradient are mediated by the Coulomb interaction \cite{Wunderlich2002,khromova2012designer}.
In the setup used here, the magnetic field gradient is $19.1$ T/m at a secular axial trap frequency of $\omega_T=2\pi\times128.4(1)\si{\kilo\hertz}$, and  $J_{12} = J_{23} = J \approx 2\pi \times \SI{31}{\Hz}$, which implies a gate time of $\pi / 1.1J \approx \SI{14.9}{\ms}$. 
The additional $J_{13} \sigma_1^z \sigma_3^z$ coupling contributes with a complex phase in the computational basis, which is irrelevant for classical computation, so we choose to omit it.
Finally, $\Omega$ is determined by the amplitude of the incident microwave radiation.
$H_{zz}$ is the Hamiltonian generating the required spin-spin interaction via magnetic gradient induced coupling (MAGIC).
Cross-talk between qubits was neglected; its main source is the non-resonant excitation of neighbouring qubits which has been measured to be on the order $10^{-5}$~ \cite{Piltz2014}.
Choosing a detuning $\delta$, such that $\omega_x = \omega_2 - \delta$, and in an appropriate rotating frame, $H_I$ reads as
\begin{equation}
    \label{eq:Hrotating}
    H_I^{(i)} \approx H_{zz}
        + \frac{\hbar \delta }{2}\sigma^z_2
        + \frac{\hbar \Omega}{2}\left( \cos(\phi)\sigma^x_2 +\sin(\phi)\sigma^y_2\right),
\end{equation}
with an error of $O(\Omega/(2\omega_2 - 2\delta))$ \cite{antonio_classical_2015}.
Choosing $\phi=0$ recovers the Hamiltonian~\eqref{eq:HTOF}.

\begin{figure*}[t]
    \subfloat[\CNOT{}]{\includegraphics[width=.33\textwidth]{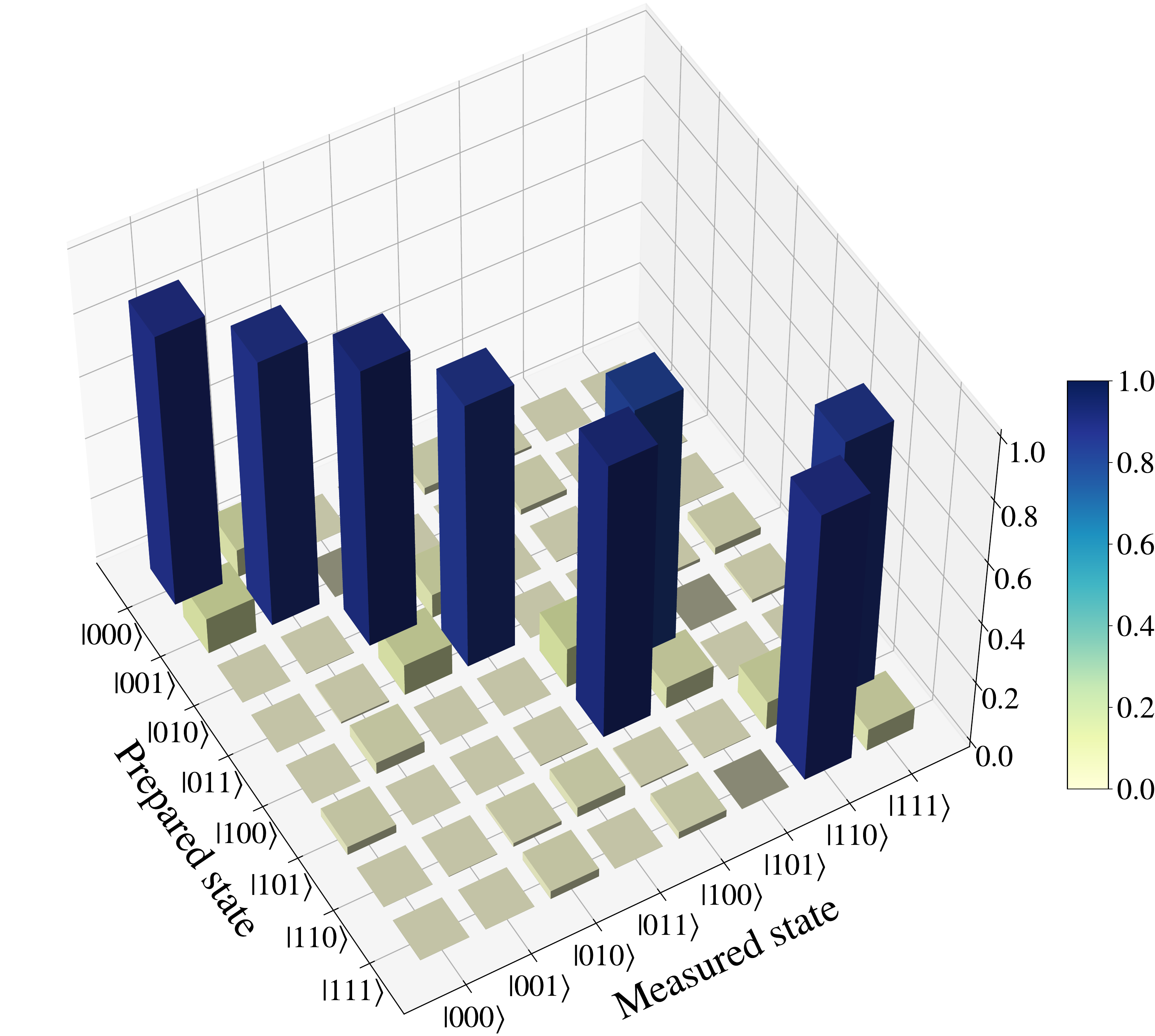}}
    \subfloat[Toffoli]{\includegraphics[width=.33\textwidth]{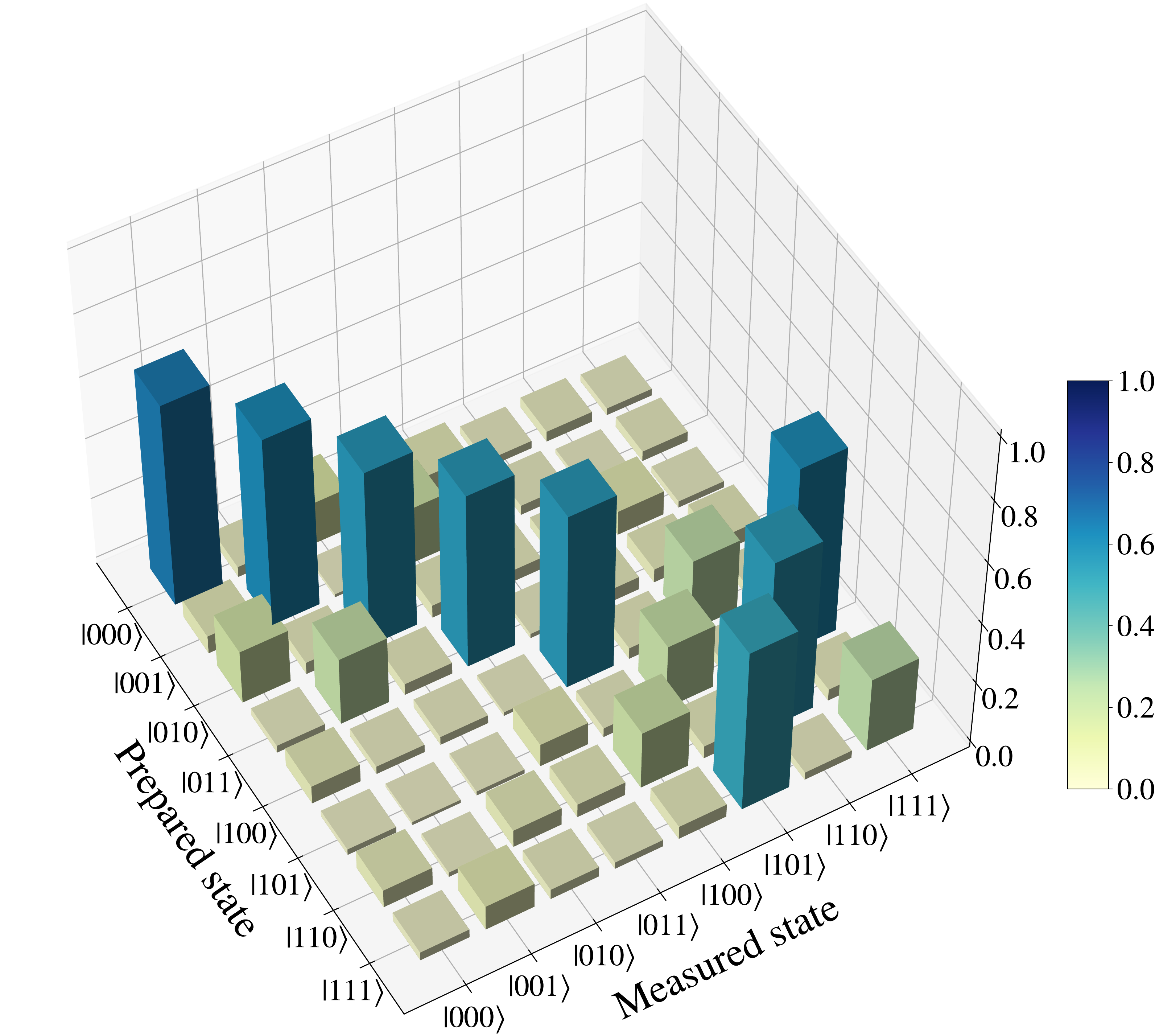}}
    \subfloat[Half-Adder]{\includegraphics[width=.33\textwidth]{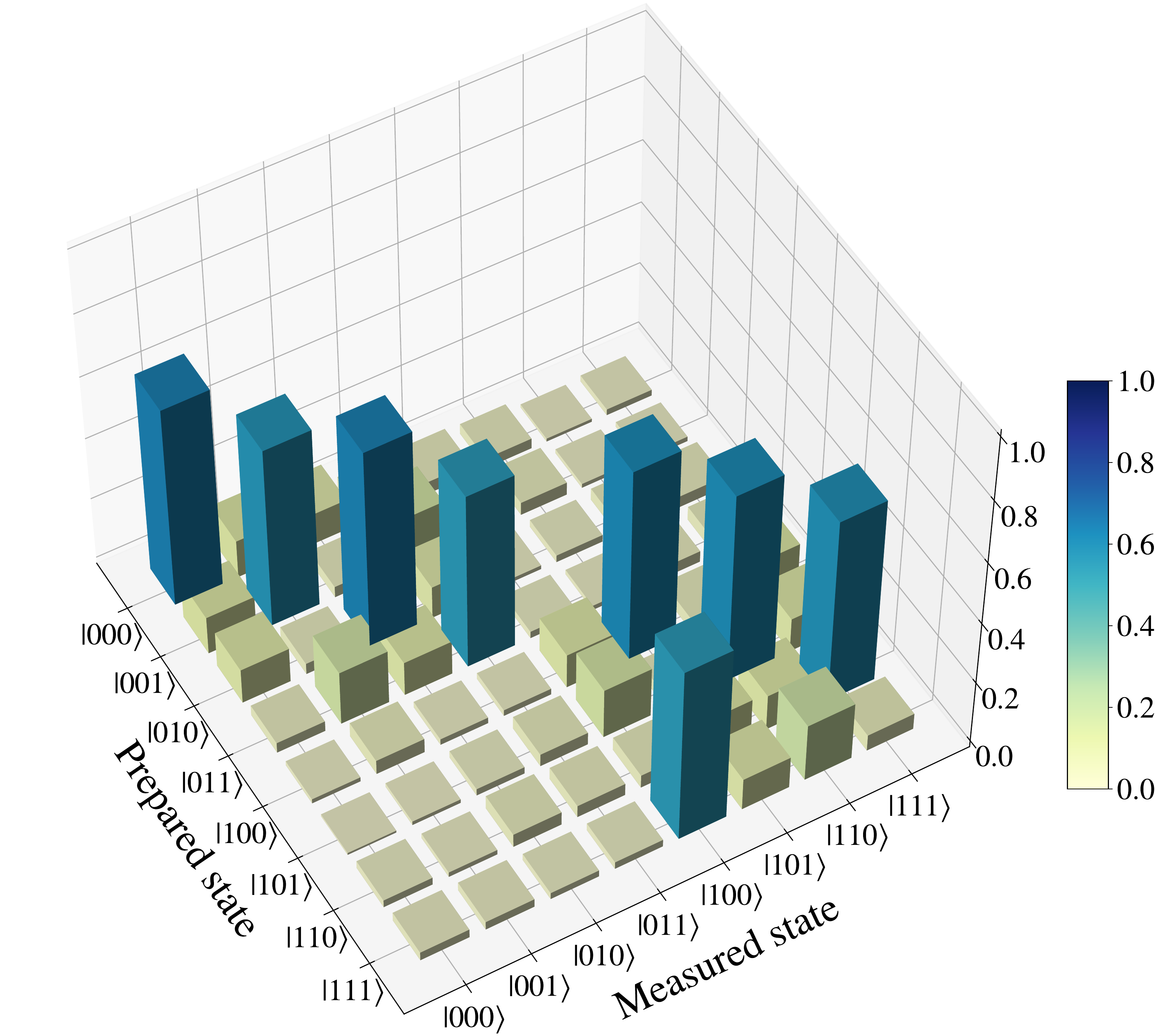}}
    \caption{Measurement probabilities for the \CNOT{}, Toffoli and Half-Adder gates, reconstructed from Maximum-Likelihood Estimate tomography. The \CNOT{} gate is controlled by qubit  1 and acts on qubit 3, while the Toffoli gate has qubit 2 as a target. The \CNOT{}, Toffoli and Half-adder gates have classical fidelities of $86.9 \%$, $58.8 \%$ and $60.6 \%$, respectively.}
    \label{fig:mle_reconstruction}
\end{figure*}

%%%%%%%%%%%%%%
% Dynamical Decoupling
%%%%%%%%%%%%%%

Fluctuations in the magnetic field dephase the qubits, which are first-order sensitive to them.
Not using passive magnetic shielding and active compensation, the  coherence time in this setup is $\approx \SI{200}{\micro\second}$ \cite{Piltz2013} -- two orders of magnitude lower than our gate times.
We thus employ Dynamical Decoupling (DD) to protect the qubits.
For  DD we intersperse single-qubit $\pi$ rotations in-between the Hamiltonian evolution, by periodically irradiating the qubits with a top hat-shaped pulse of Rabi frequency $2 \pi \times \SI{33}{\kHz}$. The $J_{ij}$ couplings are negligible during this time, since they are three orders of magnitude smaller than the Rabi frequency.
We use the notation $\theta^\phi$ for a Rabi rotation of angle $\theta$ with phase $\phi$: $\pi^0$ and $\pi^{\pi/2}$ are thus $\sigma^x$ and $\sigma^y$ gates, respectively (see Eq.~\eqref{eq:Hrotating}).
Applying a $\pi$-pulse amounts to a change of basis, so we need to change the Hamiltonian evolution accordingly. The $\sigma^z_2$ term acquires a relative minus sign, since $\sigma^z_i\sigma^{x,y}_i=-\sigma^{x,y}_i\sigma^z_i$, which we compensate by changing the microwave detuning as $\delta \rightarrow -\delta$. On the other hand, the $\sigma^z_i \sigma^z_j$ terms are left unchanged when both qubits are flipped simultaneously.
The $\sigma^x_2$ term acquires a minus sign when a $\sigma_2^y$ pulse is applied, which we compensate by adding a phase of $\phi=\pi$ to the driving field's phase.
This limits the applicable DD-sequences to either $\sigma^x$ or $\sigma^y$ pulses.
For the Toffoli gate, we choose the Carr-Purcell-Meiboom-Gill pulse sequence \cite{Maudsley1969, Piltz2013} \CPMGXY{} applied to the target qubit and a Universal Robust (\UR{}) sequence \cite{Genov2017} applied to the control qubits (see FIG.~\ref{fig:half-adder}).

The \CNOT{} gate implementation using the MAGIC scheme is realised by a $(\pi/2)^0$-pulse on the target qubit, and a unitary evolution generated by $H_{zz}$ inducing a relative phase change of $\pi$ conditioned on the logical state of the control qubit. Finally, a $(\pi/2)^{3\pi/2}$-pulse is applied to the target \cite{khromova2012designer}.  In a register exceeding size 2, it is necessary to decouple the spectator qubits from the qubits carrying out the \CNOT{} \cite{khromova2012designer}. This is achieved using a DD-sequence on the qubits participating in the \CNOT{} gate and excluding spectator qubits \cite{Piltz2013}.
The conditional evolution time used in this work is $T_\text{\CNOT}\approx \SI{8.75}{\milli\second}$. We applied a $\UR$ sequence of 120 DD pulses, with a pulse duration of $\SI{15}{\micro\second}$ each to protect the qubits coherence \cite{khromova2012quantum}.
The circuit diagram is shown in FIG.~\ref{fig:half-adder}.

%%%%%%%%%%%%%%
% Fidelities
%%%%%%%%%%%%%%

Using Maximum-Likelihood process tomography \cite{jevzek2003quantum}, we characterized the \CNOT{}, Toffoli and Half-Adder gates with classical fidelities of $F_{CNOT}=86.9 \%$, $F_T=58.8 \%$ and $F_H=60.6 \%$, respectively (see Figure~\ref{fig:mle_reconstruction}). The 90\% Confidence Intervals are respectively [0.665, 0.937], [0.221, 0.683] and [0.409, 0.662], \textit{i.e.}, the probability that the true fidelities lie in these intervals is guaranteed to be at least 90\%\cite{kiktenko2021confidence}.
The classical fidelity is the probability of obtaining the correct output given a uniformly random computational state.  
To better contextualize the fidelity we obtained for the Half-Adder, we note that a Half-Adder may alternatively be composed of single-qubit rotations and 11 \CNOT{} gates instead of taking advantage of a Toffoli gate as in this work. To then achieve the same fidelity $F_H=60.6 \%$ as obtained here with such a decomposition of the Half-Adder, and assuming nearest-neighbour couplings, each \CNOT{} would require at least a fidelity $F_{11}=95.5\%$ (see Supplemental Material). Comparing the actual \CNOT{} fidelity $F_{CNOT}=86.9 \%$ with $F_{11}=95.5\%$ shows that a lower circuit depth obtained by using a multi-qubit gate not only speeds up the Half Adder implementation but also yields a higher fidelity for given hardware parameters.

%%%%%%%%%%%%%%
% Energetics
%%%%%%%%%%%%%%

We now turn to the question of energetic cost.
In this work, we focus on the energy delivered to the Quantum Processing Unit (QPU), which we define as the physical components housed inside the vacuum chamber. The energy required for the implementation of a Half Adder is supplied to the QPU in six steps: I.\ Doppler Cooling, II.\ Sideband Cooling, III.\ State Preparation, IV.\ Toffoli and V.\ \CNOT{} (information processing), and VI.\ Readout. We are particularly interested in the cost of information processing, that is, steps IV\ and V.
All steps, except IV and V, are implemented by applying laser light near $\SI{369}{ \nano\meter}$ and near $\SI{935}{\nano\meter}$, a microwave field near 12.6 GHz, and an RF field near 20 MHz that  generates the RF trapping potential. During step IV and V, only the microwave field near 12.6 GHz is applied.  The optical power of the laser beams, the microwave power, and the power of the RF signal near 20 MHz is measured using commercial devices (more details on the experimental set-up are given in the Supplemental Material.) 

In addition to the field generating the Toffoli gate itself, there is energy necessary for the DD $\pi$-pulses. At a Rabi frequency of $2\pi\times \SI{33}{\kilo\hertz}$, microwave power of $\SI{0.58}{\watt}$ near 12.6. GHz is required in the current setup.
The Toffoli is generated using a Rabi frequency of $\SI{34}{\hertz}$, a factor $10^3$ less than the Rabi frequency used to implement single qubit $\pi$ and $\pi/2$ rotations. Due to the low power, its cost was estimated assuming the law $P \propto \Omega^2$ (see Equation~(S14) of the Supplemental Material) as $\SI{9.2}{\nano\joule}$. The power consumption of $3\times 200$ $\pi$-pulses dominates the energetic cost of the Toffoli gate.   

\begin{figure}[t] 
\centering
\resizebox{1.1\columnwidth}{!}{\includegraphics{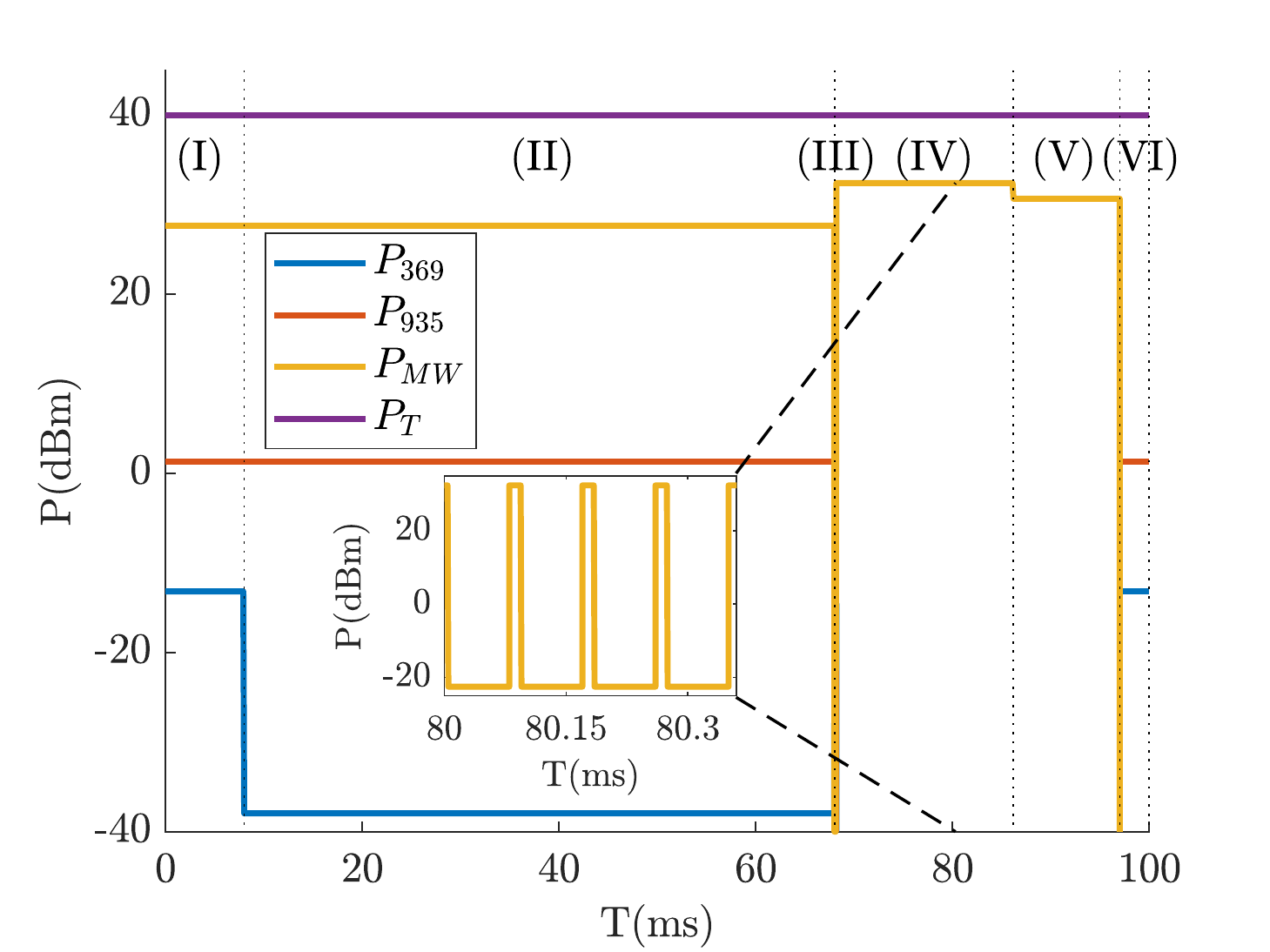}}
\caption{
Power delivered to the QPU: $P_{369}$ is the power delivered by the cooling, preparation, and readout laser; $P_{935}$, repumping laser; $P_{MW}$, microwave to close the cooling cycle and coherent qubit control; $P_T$, RF trapping potential. The time stamps are, in order, (I)  $\SI{8}{\milli\second}$ Doppler cooling, (II) $\SI{60}{\milli\second}$ sideband cooling \cite{SBC_MO}, (III) $\SI{0.2}{\milli\second}$ state preparation, (IV) $\SI{18.2}{\milli\second}$ Toffoli gate including dynamical decoupling, (V) $\SI{11}{\milli\second}$ \textsc{CNOT} gate including dynamical decoupling, and (VI) $\SI{3}{\milli\second}$ readout. In the inset, we can see a detail of the DD pulses for the Toffoli gate. See tab.~(ST1). in the supplemental material for the power measurements.}
\label{fig:power-timeline}
\end{figure}

In Table~\ref{tab:energy_estimates}, we summarise the energy estimates for all experimental steps and gates, and in FIG.~\ref{fig:power-timeline} we present the power delivered by each component as a function of time.
The table includes the number of $\pi$-pulses, equivalent to \NOT{} gates, used in the DD scheme.

\begin{table}[t]
    \input{tables/energetic_costs.tex}
    \caption{
    Measured energy consumption of the  experimental steps I - VI. The main contribution to the overall energy consumption comes from dynamical decoupling, apart from the RF trap's constant cost. 
    The energy contribution of the CNOT gate consists of two $\pi/2$-pulses totalling up to a $\pi$-pulses energy contribution. All measured energies displayed here carry a relative error of $10^{-2}$.}
    \label{tab:energy_estimates}
\end{table}

%%%%%%%%%%%%%%
% Discussion and Outlook
%%%%%%%%%%%%%%

The most efficient supercomputer today, Frontier TDS, requires $\approx \SI{e5}{\eV}$ per bit operation (see details in \cite{moutinho2022quantum}) -- 13 orders of magnitude lower than the \textsc{not} gate reported here. We emphasise that this work presents a baseline for systematically investigating energy consumption and efficiency of quantum platforms, in particular ion-trap setups. There is vast room for improvement. We exemplify this statement by showing how a next-generation ion trap will be 5 orders of magnitude more efficient.

Based on this  energetic analysis, we identify three main sources of energy consumption.
First, the microwave wave guide produces a wavefront ($\srad$ in Eq.~(S14) of the Supplemental Material) that is orders of magnitude larger than the ions' interaction cross-section. For a single \NOT{} gate, $\approx 10^{17}$ photons are irreversibly lost.
Second, the conditional gate times are much longer than the qubits' coherence times, making it necessary to use $10^3$ DD pulses, where most of the energy is spent (see Table~\ref{tab:energy_estimates}).
Third, the RF drive of the Paul trap currently produces the highest energy consumption, of the order of $\SI{3}{\watt}/\mathrm{ion},$
or $\SI{50}{\milli\joule}/\mathrm{ion}$ for the duration of a Toffoli gate.

A future planar ion trap setup \cite{Boldin2022} can address these issues. First, it integrates microwave antennae and resonators closer to the ions into a planar setup \cite{Allcock2013}, which can greatly reduce the irreversible loss of microwave photons. For example, a \NOT{} gate is performed in \SI{1.7}{\micro\second} at an applied power of \SI{10}{\milli\watt}, consuming \SI{17}{\nano\joule} of energy per gate, as opposed to $\SI{8.8}{\micro\joule}$ in this work.
In addition, the $J$-coupling, necessary for conditional gates, will be increased by about two orders of magnitude, thus reducing the time needed for \CNOT, Toffoli, and Half-Adder by the same factor. Furthermore, the coherence time is prolonged by about two orders of magnitude, mainly due to the use of magnetic field shielding. Although faster gates imply higher energy consumption (for a given geometry), the energy consumption decreases when using fewer DD pulses, or completely omitting them. We expect a Toffoli gate to require \SI{4}{\pico\joule} and a gate time of \SI{125}{\micro\second}, which eliminates the need for DD.
Implementing the \CNOT{} gate still requires two $\pi/2$ pulses on the target qubit as well as two $\pi$-pulses on the target and control to decouple them from qubit 2, resulting in 5$\pi$-pulses.
In total, \SI{85}{\nano\joule} will be required for the Half-Adder.
This is approximately $10^5$ times more efficient than the current setup (Table~\ref{tab:energy_estimates}), due to $1000\times$ more efficient pulses and $100\times$ fewer DD pulses. And still, this trap was not built with the specific goal of energy efficiency.
Finally, shorter distances between the electrodes generating the trapping potential and the ion in a planar ion trap reduce the  RF power necessary to run such a trap. Current planar traps require less than $\SI{1}{W}$ of RF power to maintain trapping.  
Furthermore, since this cost is fixed, we can increase the trap efficiency by increasing the number of ions per trap, or possibly combining traps efficiently using ion shuttling.

Cavity QED setups \cite{Haroche2020} can bring the radiation area ($\srad$) closer to the ions' effective dipole area ($\sdip$). 
In that limit, and considering the $\SI{}{\GHz}$ operation regime, Eq.~(S14) predicts a cost of on the order of \SI{}{\micro\eV} per \textsc{not} gate, which is close to the photon energy, \SI{8}{\micro\eV}.
This is 10 orders of magnitude lower than the estimate for Frontier TDS.

Efficient control protocols will become necessary in the ``one-photon limit'' \cite{banacloche-minimum}, which is still an open problem \cite{itano2003comment, banacloche-2008}.
Nevertheless, it was recently observed that, indeed, typically one quanta of energy is used from control fields\cite{stevens2021energetics}.
Other strategies to reduce energetic costs may be reusing control energy, manipulating several qubits at once and/or recycling unused energy.
Reducing the trap cost is also important, possibly using Penning traps, tighter geometries and packing more ions per trap.

In conclusion, we implemented classical logic---\NOT{}, \CNOT{}, Toffoli and Half-Adder circuits---using the quantum states of trapped ions. Our work opens the door for future implementations of classical logical on quantum technologies, with the potential energy savings of reversible computation, and presents a benchmark for future platforms.
Our work presents a path towards low energy computation beyond \textsc{cmos}, whose efficiency is decelerating.
We show
%\textcolor{blue}{
short-term steps and possible long-term paths
%}
for improvement, and hope to encourage the community to fulfil the
%\textcolor{blue}{
vision of energy-efficient computing.
%}

\section*{Supplementary Material}
Please see the supplemental material for further details on the Toffoli gate, experimental setup, power measurements and estimation and fidelity discussion.

\bigbreak

%%%%%%%%%%%%%%
% Acknowledgements
%%%%%%%%%%%%%%

We thank Lorenzo Buffoni, Marwan Mohammed and João Moutinho for helpful comments and discussions.
SP and YO thank the support from FCT, namely through projects UIDB/50008/2020 and UIDB/04540/2020. SP thanks the support from the ``la Caixa'' foundation through scholarship No. LCF/BQ/DR20/11790030.
PHH, PB, and CW acknowledge financial support from  the EU Horizon 2020 project 820314 (microQC).
Sagar Pratapsi and Patrick H. Huber contributed equally to this work.

%This article may be downloaded for personal use only. Any other use requires prior permission of the author and AIP Publishing. This article appeared in
%S.\ S.\ Pratapsi, P.\ H.\ Huber, P.\ Barthel, S.\ Bose, C.\ Wunderlich, Y.\ Omar; Classical half-adder using trapped-ion quantum bits: Toward energy-efficient computation. Appl. Phys. Lett. 9 October 2023; 123 (15): 154003.
%and may be found at https://doi.org/10.1063/5.0176719.

\section*{References}
\bibliography{references}

\end{document}

\title{
    Supplemental Material for \\
    Classical Half-Adder using Trapped-ion Quantum Bits: \\
    Towards Energy-efficient Computation
}

\author{Sagar Silva Pratapsi}%
\altaffiliation{Authors to whom correspondence should be addressed: spratapsi@tecnico.ulisboa.pt and p.huber@physik.uni-siegen.de.}
\affiliation{
Department of Physics,
Instituto Superior Técnico, University of Lisbon,
Lisbon 1049-001, Portugal.
}
\affiliation{Instituto de Telecomunicações,
Instituto Superior Técnico, University of Lisbon,
Lisbon 1049-001, Portugal.}

\author{Patrick H. Huber}
\altaffiliation{Authors to whom correspondence should be addressed: spratapsi@tecnico.ulisboa.pt and p.huber@physik.uni-siegen.de.}
\affiliation{Department of Physics, School of Science and Technology, University of Siegen, 57068 Siegen, Germany}

\author{Patrick Barthel}
\affiliation{Department of Physics, School of Science and Technology, University of Siegen, 57068 Siegen, Germany}

\author{Sougato Bose}
\affiliation{Department of Physics and Astronomy, University College London, London WC1E 6BT, UK}

\author{Christof Wunderlich}
\affiliation{Department of Physics, School of Science and Technology, University of Siegen, 57068 Siegen, Germany}

\author{Yasser Omar}
\affiliation{Center of Physics and Engineering of Advanced Materials (CeFEMA), Instituto Superior Técnico, University of Lisbon,
Lisbon 1049-001, Portugal}
\affiliation{PQI -- Portuguese Quantum Institute,
Lisbon 1600-531, Portugal}

\date{\today}

\maketitle

\renewcommand{\theequation}{S\arabic{equation}}
\renewcommand{\thefigure}{S\arabic{figure}} 
\renewcommand{\thetable}{S\arabic{table}} 

\section{Choosing $\Omega$ coupling and gate time for the Toffoli evolution}
\label{annex:gate-ratios}
We simulated the Hamiltonian evolution
\begin{equation}
    \label{eq:HTOF}
    H_\text{TOF} =
        \frac{\hbar J}{2} \left(\sigma^z_1 \sigma^z_2 + \sigma^z_2 \sigma^z_3 \right)
        + \frac{\hbar \delta }{2}\sigma^z_2
        + \frac{\hbar \Omega}{2} \sigma^x_2
\end{equation}
numerically, for different values of $\Omega$ and a total evolution time of $\pi / \Omega$. Given our limited coherence time, we focused on finding the fastest possible Toffoli gate.

\begin{figure}[th]
    \resizebox{\columnwidth}{!}{\input{figures/gate_ratios.pgf}}
    \caption{
    \textbf{Classical Toffoli gate fidelity} as a function of the $\Omega$ coupling's strength in Eq.~\eqref{eq:HTOF}, measured in units of $J=J_{12}=J_{23}$. We simulated the Hamiltonian for a time of $\pi / \Omega$ (bottom axis). The right-most peak, with $\Omega = 1.1 J$, allows for a $\SI{14.9}{\milli\second}$ gate with approximately $99\%$ fidelity.}
    \label{fig:toffoli-fidelity}
\end{figure}
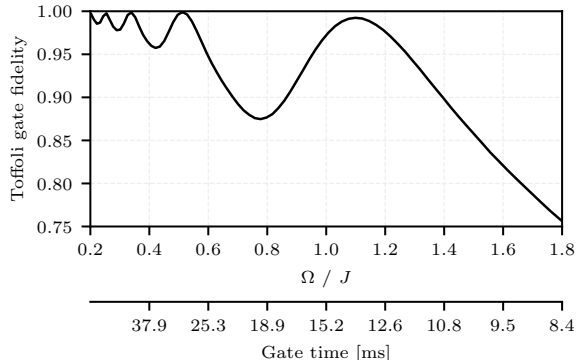

We calculated the classical fidelity of each gate as the probability of obtaining the correct classical output state when given a uniformly random input computational state, when compared to a Toffoli gate. Our findings are summarised in FIG.~\ref{fig:toffoli-fidelity}.

We chose the fastest possible gate time, with $\approx \SI{14.9}{\ms}$, corresponding to the right-most peak in FIG.~\ref{fig:toffoli-fidelity}. It is possible to find gate times with better fidelities. Although we are currently limited by coherence time, stronger $J$ couplings might allow the other peaks of FIG.~\ref{fig:toffoli-fidelity} to be experimentally realisable in the future.

\section{Theoretical energy estimate}
\label{annex:energetics}
The Rabi frequency $\Omega_R$ of a magnetic dipolar transition, like that between the two states $\ket 0$ and $\ket 1$, is a function of the magnetic field surrounding the ion,
\begin{equation}
    \hbar \Omega_R = \bm \mu \cdot \bm B_\text{amp},
    \label{eq:rabi-magnetic-field}
\end{equation}
where $\bm B_\text{amp}$ is the magnetic field amplitude and $\bm{\mu} = \mel{1}{\bm M}{0}$ is the magnetic dipole moment of the transition. The Rabi frequency $\Omega_R$ is experimentally accessible.
We estimate $\mu = \mu_B,$
where $\mu_B$ is the Bohr magneton. Having these two quantities, we can estimate the value of $\bm B_\text{amp}$ which, in turn, will allow us to calculate the energy carried by the electromagnetic field of the microwave pulse. To do so, we need to know the profile of the microwave field's wave front.

To generate the microwave field, we use an OFHC copper cylindrical cavity resonator of diameter $2a \approx \SI{16.3}{\milli\meter}$, placed inside the vacuum chamber \cite{khromova2012quantum}.
The dimensions of the cavity were chosen specifically to allow the propagation of the dominant transverse-electric mode, \TE{}, at the qubit frequency $\SI{12.6}{\GHz}$, while suppressing higher-order modes.

To estimate the energy carried by the microwave's field, we assume that the ions are placed near the open end of the microwave cavity, close to its cylindrical axis, and that the magnetic field at that point may be approximated by the \TE{} mode of an infinite circular wave guide of the same diameter. Let us use a cylindrical coordinate system $\rho, \phi, z$, where $z$ is aligned with the cylinder axis. Let us calculate the intensity of the fields of the \TE{} mode by assuming a time and  $z$ dependence like
\begin{equation}
    \bm E, \bm B \propto e^{i(\omega t -\beta z)}.
    \label{eq:z-dependence}
\end{equation}

From Maxwell's equations for the vacuum inside the waveguide we can obtain expressions for the EM fields as a function of the $E_z$ and $B_z$ components
\begin{align}
    \begin{bmatrix}
        E_\rho \\
        B_\phi
    \end{bmatrix}
    & =
    \frac{-i}{k_c^2}
    \begin{bmatrix}
        \beta             & \phantom{+} \omega \\
        \phantom{+} k/c   & \beta
    \end{bmatrix}
    \begin{bmatrix}
        \partial_\rho E_z \phantom{/ \rho} \\
        \partial_\phi B_z / \rho
    \end{bmatrix}
    \nonumber
    \\
    \begin{bmatrix}
        B_\rho \\
        E_\phi
    \end{bmatrix}
    & =
    \frac{-i}{k_c^2}
    \begin{bmatrix}
        \beta & -k/c \\
        -\omega  & \beta
    \end{bmatrix}
    \begin{bmatrix}
        \partial_\rho B_z \phantom{/ \rho} \\
        \partial_\phi E_z / \rho
    \end{bmatrix},
    \label{eq:maxwell}
\end{align}
where $k_c^2 = k^2 - \beta^2$ and $k = \omega / c.$

The resonator was designed to allow only transverse electric fields, that is, $E_z=0$.
Applying $\bm \nabla \cdot \bm B = 0$, and assuming a separation of variables $B_z = B_0 R(\rho) \Phi(\phi)$, we arrive at
\begin{align}
    \partial_\phi^2 \Phi + \alpha^2 &= 0
    \label{eq:phi-component} \\
    \left(\rho \partial_\rho\right)^2 R + (\rho^2 k_c^2 - \alpha^2)R &= 0,
    \label{eq:bessel}
\end{align}
for some constant $\alpha$.

Equation~\eqref{eq:phi-component} has the general solution $\Phi(\phi) = A \cos\left(\alpha(\phi - \phi_0)\right)$.
Imposing $\Phi(\phi + 2\pi) = \Phi(\phi)$ restricts $\alpha$ to integer values.
Equation~\eqref{eq:bessel} is the \textit{Bessel functional equation.} The only solutions that do not diverge are the Bessel functions of the first kind, $J_\alpha(k_c \rho)$. Since $J_{-\alpha}$ and $J_\alpha$ are equal up to a $\pm 1$ factor, we will take $\alpha$ to be a non-negative integer and rename it to $\alpha = n \in \mathbb N$.
We thus arrive at the solution
\begin{equation}
    B_z = B_0 \, J_n(k_c \rho) \, \cos(n(\phi - \phi_0)) \, e^{i(\omega t - \beta z)}.
\end{equation}

The remaining field components can be calculated from Equations~\eqref{eq:maxwell},
\begin{align*}
\begin{split}
    E_\rho &= \phantom{+} \frac{i\omega B_0 n}{k_c}           \, \frac{J_n(k_c \rho)}{k_c\rho} \, \sin(n(\phi - \phi_0)) \, e^{i(\omega t-\beta z)} \\
    E_\phi &= \phantom{+} \frac{i\omega B_0}{k_c} \phantom{n} \, J_n'(k_c \rho)                \, \cos(n(\phi - \phi_0)) \, e^{i(\omega t-\beta z)} \\
    B_\rho &=          -  \frac{i\beta B_0}{k_c}  \phantom{n} \, J_n'(k_c \rho)                \, \cos(n(\phi - \phi_0)) \, e^{i(\omega t-\beta z)} \\
    B_\phi &= \phantom{+} \frac{i\beta B_0 n}{k_c}            \, \frac{J_n(k_c \rho)}{k_c\rho} \, \sin(n(\phi - \phi_0)) \, e^{i(\omega t-\beta z)},
    \label{eq:explicit_fields}
\end{split}
\end{align*}
where $J'_n$ is the derivative of the $n$-th Bessel function.

Finally, there is the added restriction that $\bm E$ be perpendicular to the conductor's surface, that is, $E_\phi(\rho = a) = 0$, which means that
\begin{equation}
    k_c a = p_{nm}',
\end{equation}
where $p_{nm}'$ is the $m$-th root of $J_n'$. For each $n, m$ there is a different solution -- the so-called TE\textsubscript{nm} modes.

We will now focus on the TE\textsubscript{11} mode, for $n=1$, which is the first non-zero solution to Equations~\eqref{eq:phi-component} and \eqref{eq:bessel}, also called the \textit{dominant mode}, and also the only allowed mode of propagation by our cavity.
In this case, we have $k_c a = p'_{11} \approx 1.8412.$

To calculate the energy carried by the microwave pulse, we calculate the Poynting vector and average it over a period $T = 2\pi / \omega$,
\begin{equation}
    \bm S^\text{avg} =
        \frac{1}{T}
        \int_0^{T}
        \frac{ \bm E \times \bm B }{\mu_0}
        \, \diff t.
\end{equation}
The only non-zero component of this vector is in the $z$ direction,
\begin{align}
\begin{split}
    S^\text{avg}_z =
    \frac 1 2
        \frac{B_0^2}{\mu_0}
        \frac{\omega \beta}{k_c^2}
        \Bigg[
            \Bigg(
                \frac{J_1(k_c \rho)}{k_c \rho}
                & \sin(\omega t - \beta z)
            \Bigg)^2
            \\
            +
            \Big(
                J_1'(k_c \rho)
                & \cos(\omega t - \beta z)
            \Big)^2
        \Bigg].
\end{split}
\end{align}

We now integrate over the entire cross-section of the cylinder to obtain the power delivered by the cavity,
\begin{equation}
    P = 
        \frac 1 2 \,
        \frac{B_0^2}{\mu_0} \,
        \frac{\omega \beta}{k_c^2} \,
        \frac{I_{11}}{(p_{11}')^2} \,
        \pi a^2,
\end{equation}
where $I_{11}$ is an integral of Bessel functions,
\begin{equation}
    I_{11} = \int_0^{p_{11}'}
        \Bigg[
            \Bigg(
                \frac{J_1(r)}{r}
            \Bigg)^2
            +
            \Big(
                J_1'(r)
            \Big)^2
        \Bigg]
        \, r \, \diff r
        \approx 0.405.
        \label{eq:I11}
\end{equation}

Finally, to estimate the power, we need only find the value of $B_0$ that makes the qubit states flop at the observed Rabi frequency $\Omega_R$.  The magnetic field amplitude at the center of the waveguide is
\begin{equation}
    B_\text{amp}
    = \lVert \bm B(\rho=0) \rVert_\text{max}
    = \frac{\beta}{k_c} \frac{B_0}{2}
    = \frac{\hbar \Omega_R}{\mu\, \cos \theta}.
\end{equation}
The last equality comes from Equation~\eqref{eq:rabi-magnetic-field}.
The angle $\theta$ is that between the ions' magnetic dipole moment and the cavity's magnetic field.

We then arrive at the expression for the power carried by the microwave pulse,
\begin{equation}
    P =
        \frac 1 2
        \frac{\srad}{\sdip}
        \frac{\hbar \Omega_R^2}{\cos^2 \theta},
\end{equation}
where we defined the areas
\begin{equation}
    \srad =
        \frac{4 I_{11}}{(p'_{11})^2}
        \frac{\pi a^2}{\sqrt{1 - x^2}}
    \quad\text{and}\quad
    \sdip =
        \frac{\mu_0}{\hbar c}
        \mu^2,
\end{equation}
with $x = c p'_{11} / \omega a$. The area $\sdip$ can be thought of as an effective dipole cross-section for the ion.

%\color{blue}

\section{Experimental energy measurements}
\subsection{Experimental component schematics}
\begin{figure}[t] 
\centering
\includegraphics[width=.9\textwidth]{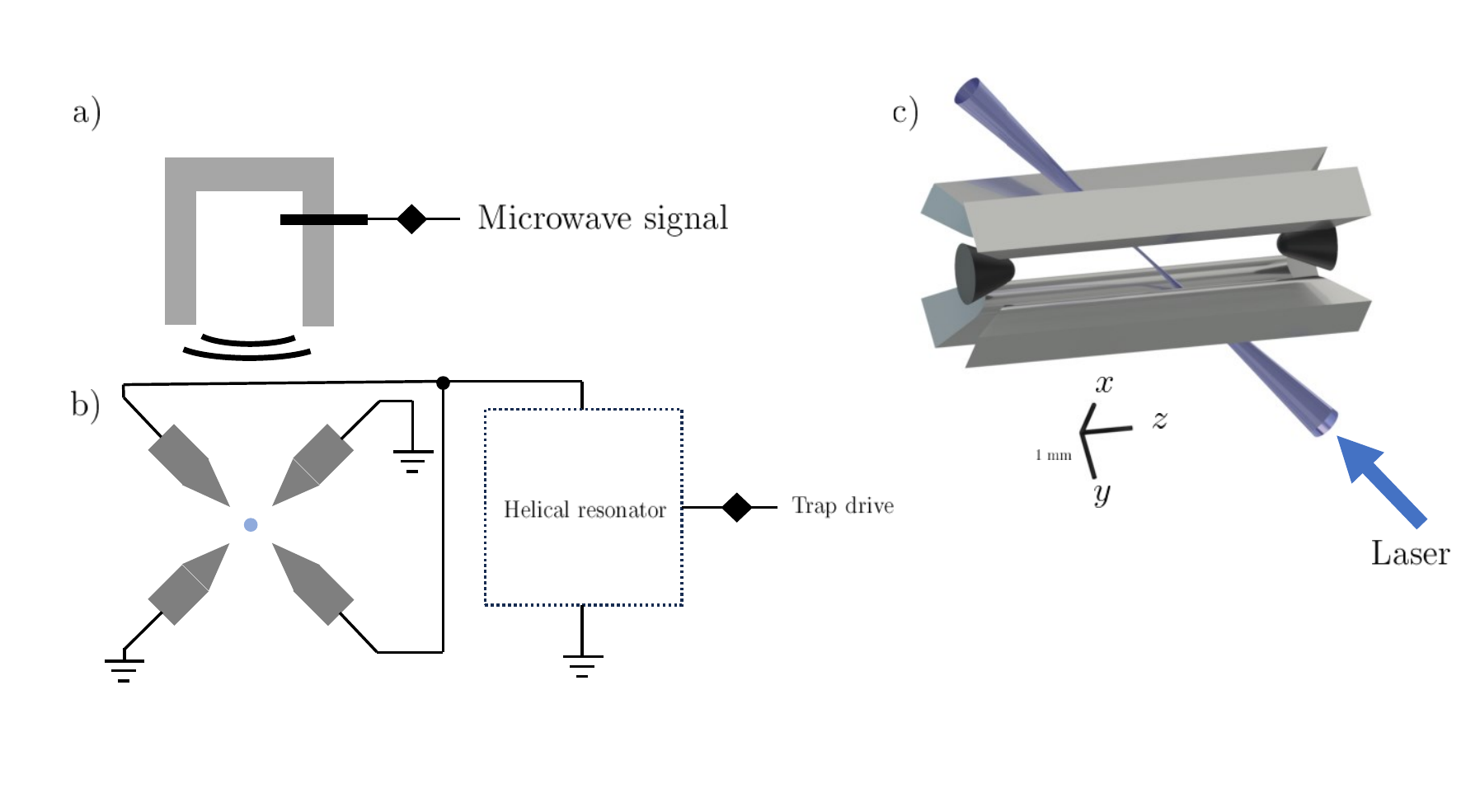}
\caption{
%\color{blue}
a) Circular wave guide to provide the microwave control signal for coherent qubit manipulation and cooling.
b) Schematic of RF voltage supply of the trap to generate the radial confinement in the linear Paul-trap. For galvanic isolation and noise filtering, the trap drive is supplied to the experimental setup using a helical resonator. 
c) Global laser field applied to the Ytterbium ions containing laser light of wavelength $\SI{369}{nm}$, and $\SI{935}{nm}$ light for cooling preparation readout and optical pumping. Diamonds indicate where a power meter was placed to measure the power delivered to the respective device.} 
\label{fig:schematics}
\end{figure}

As depicted in FIG.~\ref{fig:schematics}, energy is supplied to the QPU by the  microwave signal near 12.6 GHz for quantum control and cooling and the laser field required for cooling, state initialisation, and read-out. Table ~\ref{tab:one-time-costs} shows the contributions from laser light and microwaves, respectively. The RMS RF power near 20 MHz applied to the resonator is  $\SI{11\pm 1}{\watt}$. 
\begin{table}[t]
\begin{tabular}{@{}lrrrrr@{}}
\toprule
\textbf{}                                                      & \multicolumn{1}{c}{{\begin{tabular}[c]{@{}c@{}}Laser \\ 369 nm\end{tabular}}} & \multicolumn{1}{c}{{\begin{tabular}[c]{@{}c@{}}Laser \\ 935 nm\end{tabular}}} & \multicolumn{1}{c}{{MW/ion}} & \multicolumn{1}{c}{{Duration}}  & \multicolumn{1}{c}{{Energy}} \\ \midrule
\begin{tabular}[c]{@{}l@{}}Doppler\\ cooling\end{tabular}          & $\SI{48.0}{\micro\watt}$  & $\SI{1.35}{\milli\watt}$  & $\SI{0.58}{\watt}$      & $\SI{8.0}{\milli\second}$  & $\SI{14}{\milli\joule}$           \\
\begin{tabular}[c]{@{}l@{}}Sideband\\ cooling\end{tabular}         & $\SI{0.16}{\micro\watt}$  & $\SI{1.35}{\milli\watt}$  & $\SI{0.58}{\watt}$      & $\SI{60}{\milli\second}$   & $\SI{100}{\milli\joule}$          \\
\begin{tabular}[c]{@{}l@{}}Ground \\ state prep. \end{tabular}     & $\SI{35.0}{\micro\watt}$  & $\SI{1.35}{\milli\watt}$  & \multicolumn{1}{c}{---} & $\SI{0.20}{\milli\second}$ & $\SI{0.28}{\micro\joule}$         \\
Readout                                                            & $\SI{48.0}{\micro\watt}$  & $\SI{1.35}{\milli\watt}$  & \multicolumn{1}{c}{---} & $\SI{3.0}{\milli\second}$  & $\SI{4.2}{\micro\joule}$          \\ \midrule
\textbf{}                                                          &                           &                           &                         & {Total}             & $\SI{120}{\milli\joule}$          \\ \bottomrule
\end{tabular}
\caption{
%\color{blue}
Power and energy costs of ``one-time'' operations that contribute to the baseline energy expenditure. MW: Microwave. The measured powers have a relative error of $10^{-2}$.
}
\label{tab:one-time-costs}
\end{table}

\quad \\
\onecolumngrid

\twocolumngrid 
\quad \\

\section{Equivalent \textsc{cnot} fidelity from Half-Adder decomposition}
\begin{figure}[h] 

\begin{tikzpicture}
\node[scale=0.7]{
\begin{quantikz}
\ket{a}\quad\qw&\qw& \qw&\qw&\ctrl{1}&\qw&\qw&\qw&\ctrl{1}&\qw&\qw&																												\ctrl{1}&\targ{}		&\ctrl{1}	&\qw				&\qw							&\qw \slice{}
&\qw&\ctrl{1}&\targ{}&\ctrl{1}&\qw\\
\ket{0}\quad \qw& \gate{H} & \targ{}&\gate{T^\dagger}&\targ{}&\gate{T}&\targ{}&\gate{T^\dagger}&\targ{}&\gate{T}&\gate{H}&	\targ{}	&\ctrl{-1}	&\targ{}	&\ctrl{1}		&\gate{T}					&\ctrl{1}
&\ctrl{1}&\targ{}&\ctrl{-1}&\targ&\qw&\qw\\
\ket{c}\quad \qw&\qw&\ctrl{-1}&\qw&\qw&\qw&\ctrl{-1}&\qw&\qw&\gate{T}&\qw&																								\qw			&\qw				&\qw			&\targ{}		&\gate{T^\dagger}	&\targ{}
&\targ{}&\qw&\qw&\qw&\qw
\end{quantikz}};
\end{tikzpicture}
\caption{
Decomposition of the Half-Adder circuit into next neighbour \textsc{cnot} operations and local rotations. Here, $T=Z^{1/4}$  and $H$ is the Hadamard gate. The vertical line separates the Toffoli from the \textsc{cnot} operation realised by a swap gate and a next-neighbour \textsc{cnot} gate to implement the Half-Adder circuit. Each \textsc{cnot} operation would require a conditional evolution time $T_C=\SI{6.84}{ms}$ totalling up to $T=\SI{89}{ms}$ in the three-qubit system used to demonstrate the direct implementation of the Toffoli and the Half-Adder gate. The \textsc{cnot} gates on both sides of the vertical line cancel out, which we take into account in our noise model.
}  
\label{fig:half-adder-decomposition}
\end{figure}
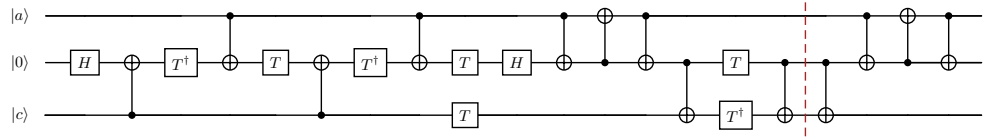

Quantum platforms that do not have three-qubit gates may implement a Half-Adder gate by decomposing it into a combination of single- and two-qubit gates. In FIG.~\ref{fig:half-adder-decomposition}, we show such a decomposition using \textsc{cnot} and single-qubit rotation gates.
We may then ask what is the \textsc{cnot} fidelity required to achieve the Half-Adder fidelity we obtained experimentally.

To estimate this number, we will consider a depolarising noise model. To do so, consider the Pauli Transfer Matrix (PTM) \cite{greenbaum2015introduction} $P$ associated with a given Completely Positive and Trace-Preserving (CPTP) $n$-qubit quantum operation $\Lambda$, that is,
\[
    [P_{ij}] = [\tr{P_i \Lambda(P_j)}] = 
    \begin{pmatrix}
        1 & 0 \\
        v & A
    \end{pmatrix},
\]
with $P_i$ being the $i$-th Pauli string in $\{\sigma^0, \sigma^x, \sigma^y, \sigma^z\}^{\otimes n}$, using lexicographical ordering. Here, $A$ is an $m \times m$ matrix, $v$ is a column vector of dimension $m$ and $0$ is a row vector of $m$ zeroes, with $m=4^n - 1.$

In the PTM formalism, a depolarising channel has the simple representation  \cite{greenbaum2015introduction}
\[
P_\mathrm{depolarising} = 
\begin{pmatrix}
    1 & 0 \\
    0 & \lambda I 
\end{pmatrix},
\]
where $0 \leq \lambda \leq 1$ and $I$ is the $m \times m$ identity matrix. Given that PTM matrices are multiplicative -- that is, the PTM matrix of the composition of two CPTP maps is just the product of their respective PTM matrices -- we will apply a depolarising noise to a unitary process by multiplying its $A$ submatrix by a factor $\lambda$, resulting in a new PTM matrix
\[
P'=
    \begin{pmatrix}
        1 & 0 \\
        v & \lambda A
    \end{pmatrix}.
\]
It is worth pointing out that $v=0$ for unitary processes like the $\textsc{cnot}.$ We also note that, if a given unitary applies only to $q$ qubits, we only apply a depolarising channel to those $q$ qubits; the spectator qubits will remain unaffected.

Following the procedure just described, we assume depolarising noise characterized by $\lambda = 0.999$ to the Hadamard gates, in line with state-of-the-art ion trap platforms.
On the other hand, the $T=Z^{1/4}$ (and $T^\dagger$) gates will remain noiseless, as they are usually virtual gates, achieved via a phase shift of the incident driving field, usually a microwave or laser field. Numerically, we conclude that, to achieve a Half-Adder with $60.6\%$ fidelity following the circuit in FIG.~\ref{fig:half-adder-decomposition}, the \textsc{cnot} gates must have a fidelity of at least 95.5\%.

\onecolumngrid

\twocolumngrid

\quad
\newpage
\pagebreak
\bibliographystyle{naturemag}
\bibliography{references}

%% file: tables/energetic_costs.tex
\begin{tabular}{@{}lccccrr@{}}
\toprule
    \multirow{2}{*}{\textbf{Operation}}
    & \multirow{3}{*}{\pbox{1cm}{\textbf{Laser} \SI{369}{\nano\meter}}}
    & \multirow{3}{*}{\pbox{1cm}{\textbf{Laser} \SI{935}{\nano\meter}}}
    & \multicolumn{3}{c}{\textbf{Microwave}}
    & \multirow{3}{*}{\textbf{Total}} \\
    \cmidrule(lr){4-6}
    &
    &
    & \multirow{2}{*}{\textbf{Pulse}}
    & \multicolumn{2}{c}{\textbf{Dyn. decoupling}} \\
    \cmidrule(lr){5-6}
    &
    &
    &
    & \# $\pi$-pulses
    & Cost
    & \\
    &
    &
    &
    & (\SI{8.8}{\micro\joule})
    & 
    & \\
\midrule
\textbf{I.} Doppler c.    & $\SI{380}{\nano\joule}$ & $\SI{11}{\micro\joule}$  & $\SI{4.6}{\milli\joule}$ & --- & --- & \SI{4.6}{\milli\joule} \\
\textbf{II.} Sideband c.   & $\SI{10}{\nano\joule}$  & $\SI{81}{\micro\joule}$  & $\SI{35}{\milli\joule}$  & --- & --- & \SI{35}{\milli\joule} \\
\textbf{III.} State prep. & $\SI{7}{\nano\joule}$   & $\SI{0.3}{\micro\joule}$ & \multicolumn{1}{c}{---}  & --- & --- &  \SI{0.3}{\micro\joule} \\
%$\pi$-pulse        & --- &  --- & \SI{8.8}{\micro\joule} & ---          & \multicolumn{1}{c}{---} & \SI{8.8}{\micro\joule}  \\
\textbf{IV.} Toffoli    & --- &  --- & \SI{9.2}{\nano\joule}  & $3\times200$ & \SI{1.1}{\milli\joule}  & \SI{1.1}{\milli\joule}  \\
\textbf{V.} CNOT       & --- &  --- & \SI{8.8}{\micro\joule} & $2\times120$ & \SI{0.44}{\milli\joule} & \SI{0.44}{\milli\joule} \\
\textbf{VI.} Readout & $\SI{140}{\nano\joule}$   & $\SI{4}{\micro\joule}$ & \multicolumn{1}{c}{---}  & --- & --- & \SI{4.2}{\micro\joule} \\
\bottomrule
 \\
 \bottomrule
\multicolumn{3}{l}{Half-Adder (\textbf{IV.} + \textbf{V.})} &  &      &   & \SI{1.5}{\milli\joule} \\
\bottomrule
\end{tabular}

%% file: figures/gate_ratios.pgf
%% Creator: Matplotlib, PGF backend
%%
%% To include the figure in your LaTeX document, write
%%   \input{<filename>.pgf}
%%
%% Make sure the required packages are loaded in your preamble
%%   \usepackage{pgf}
%%
%% Also ensure that all the required font packages are loaded; for instance,
%% the lmodern package is sometimes necessary when using math font.
%%   \usepackage{lmodern}
%%
%% Figures using additional raster images can only be included by \input if
%% they are in the same directory as the main LaTeX file. For loading figures
%% from other directories you can use the `import` package
%%   \usepackage{import}
%%
%% and then include the figures with
%%   \import{<path to file>}{<filename>.pgf}
%%
%% Matplotlib used the following preamble
%%
\begingroup%
\makeatletter%
\begin{pgfpicture}%
\pgfpathrectangle{\pgfpointorigin}{\pgfqpoint{3.200000in}{2.000000in}}%
\pgfusepath{use as bounding box, clip}%
\begin{pgfscope}%
\pgfsetbuttcap%
\pgfsetmiterjoin%
\pgfsetlinewidth{0.000000pt}%
\definecolor{currentstroke}{rgb}{1.000000,1.000000,1.000000}%
\pgfsetstrokecolor{currentstroke}%
\pgfsetstrokeopacity{0.000000}%
\pgfsetdash{}{0pt}%
\pgfpathmoveto{\pgfqpoint{0.000000in}{0.000000in}}%
\pgfpathlineto{\pgfqpoint{3.200000in}{0.000000in}}%
\pgfpathlineto{\pgfqpoint{3.200000in}{2.000000in}}%
\pgfpathlineto{\pgfqpoint{0.000000in}{2.000000in}}%
\pgfpathlineto{\pgfqpoint{0.000000in}{0.000000in}}%
\pgfpathclose%
\pgfusepath{}%
\end{pgfscope}%
\begin{pgfscope}%
\pgfsetbuttcap%
\pgfsetmiterjoin%
\definecolor{currentfill}{rgb}{1.000000,1.000000,1.000000}%
\pgfsetfillcolor{currentfill}%
\pgfsetlinewidth{0.000000pt}%
\definecolor{currentstroke}{rgb}{0.000000,0.000000,0.000000}%
\pgfsetstrokecolor{currentstroke}%
\pgfsetstrokeopacity{0.000000}%
\pgfsetdash{}{0pt}%
\pgfpathmoveto{\pgfqpoint{0.479942in}{0.734383in}}%
\pgfpathlineto{\pgfqpoint{3.086474in}{0.734383in}}%
\pgfpathlineto{\pgfqpoint{3.086474in}{1.924572in}}%
\pgfpathlineto{\pgfqpoint{0.479942in}{1.924572in}}%
\pgfpathlineto{\pgfqpoint{0.479942in}{0.734383in}}%
\pgfpathclose%
\pgfusepath{fill}%
\end{pgfscope}%
\begin{pgfscope}%
\pgfsetbuttcap%
\pgfsetmiterjoin%
\definecolor{currentfill}{rgb}{1.000000,1.000000,1.000000}%
\pgfsetfillcolor{currentfill}%
\pgfsetlinewidth{0.000000pt}%
\definecolor{currentstroke}{rgb}{0.000000,0.000000,0.000000}%
\pgfsetstrokecolor{currentstroke}%
\pgfsetstrokeopacity{0.000000}%
\pgfsetdash{}{0pt}%
\pgfpathmoveto{\pgfqpoint{0.479942in}{0.317817in}}%
\pgfpathlineto{\pgfqpoint{3.086474in}{0.317817in}}%
\pgfpathlineto{\pgfqpoint{3.086474in}{0.317817in}}%
\pgfpathlineto{\pgfqpoint{0.479942in}{0.317817in}}%
\pgfpathlineto{\pgfqpoint{0.479942in}{0.317817in}}%
\pgfpathclose%
\pgfusepath{fill}%
\end{pgfscope}%
\begin{pgfscope}%
\pgfsetbuttcap%
\pgfsetroundjoin%
\definecolor{currentfill}{rgb}{0.000000,0.000000,0.000000}%
\pgfsetfillcolor{currentfill}%
\pgfsetlinewidth{0.803000pt}%
\definecolor{currentstroke}{rgb}{0.000000,0.000000,0.000000}%
\pgfsetstrokecolor{currentstroke}%
\pgfsetdash{}{0pt}%
\pgfsys@defobject{currentmarker}{\pgfqpoint{0.000000in}{-0.048611in}}{\pgfqpoint{0.000000in}{0.000000in}}{%
\pgfpathmoveto{\pgfqpoint{0.000000in}{0.000000in}}%
\pgfpathlineto{\pgfqpoint{0.000000in}{-0.048611in}}%
\pgfusepath{stroke,fill}%
}%
\begin{pgfscope}%
\pgfsys@transformshift{0.805759in}{0.317817in}%
\pgfsys@useobject{currentmarker}{}%
\end{pgfscope}%
\end{pgfscope}%
\begin{pgfscope}%
\definecolor{textcolor}{rgb}{0.000000,0.000000,0.000000}%
\pgfsetstrokecolor{textcolor}%
\pgfsetfillcolor{textcolor}%
\pgftext[x=0.805759in,y=0.220595in,,top]{\color{textcolor}\rmfamily\fontsize{7.000000}{8.400000}\selectfont 37.9}%
\end{pgfscope}%
\begin{pgfscope}%
\pgfsetbuttcap%
\pgfsetroundjoin%
\definecolor{currentfill}{rgb}{0.000000,0.000000,0.000000}%
\pgfsetfillcolor{currentfill}%
\pgfsetlinewidth{0.803000pt}%
\definecolor{currentstroke}{rgb}{0.000000,0.000000,0.000000}%
\pgfsetstrokecolor{currentstroke}%
\pgfsetdash{}{0pt}%
\pgfsys@defobject{currentmarker}{\pgfqpoint{0.000000in}{-0.048611in}}{\pgfqpoint{0.000000in}{0.000000in}}{%
\pgfpathmoveto{\pgfqpoint{0.000000in}{0.000000in}}%
\pgfpathlineto{\pgfqpoint{0.000000in}{-0.048611in}}%
\pgfusepath{stroke,fill}%
}%
\begin{pgfscope}%
\pgfsys@transformshift{1.131575in}{0.317817in}%
\pgfsys@useobject{currentmarker}{}%
\end{pgfscope}%
\end{pgfscope}%
\begin{pgfscope}%
\definecolor{textcolor}{rgb}{0.000000,0.000000,0.000000}%
\pgfsetstrokecolor{textcolor}%
\pgfsetfillcolor{textcolor}%
\pgftext[x=1.131575in,y=0.220595in,,top]{\color{textcolor}\rmfamily\fontsize{7.000000}{8.400000}\selectfont 25.3}%
\end{pgfscope}%
\begin{pgfscope}%
\pgfsetbuttcap%
\pgfsetroundjoin%
\definecolor{currentfill}{rgb}{0.000000,0.000000,0.000000}%
\pgfsetfillcolor{currentfill}%
\pgfsetlinewidth{0.803000pt}%
\definecolor{currentstroke}{rgb}{0.000000,0.000000,0.000000}%
\pgfsetstrokecolor{currentstroke}%
\pgfsetdash{}{0pt}%
\pgfsys@defobject{currentmarker}{\pgfqpoint{0.000000in}{-0.048611in}}{\pgfqpoint{0.000000in}{0.000000in}}{%
\pgfpathmoveto{\pgfqpoint{0.000000in}{0.000000in}}%
\pgfpathlineto{\pgfqpoint{0.000000in}{-0.048611in}}%
\pgfusepath{stroke,fill}%
}%
\begin{pgfscope}%
\pgfsys@transformshift{1.457392in}{0.317817in}%
\pgfsys@useobject{currentmarker}{}%
\end{pgfscope}%
\end{pgfscope}%
\begin{pgfscope}%
\definecolor{textcolor}{rgb}{0.000000,0.000000,0.000000}%
\pgfsetstrokecolor{textcolor}%
\pgfsetfillcolor{textcolor}%
\pgftext[x=1.457392in,y=0.220595in,,top]{\color{textcolor}\rmfamily\fontsize{7.000000}{8.400000}\selectfont 18.9}%
\end{pgfscope}%
\begin{pgfscope}%
\pgfsetbuttcap%
\pgfsetroundjoin%
\definecolor{currentfill}{rgb}{0.000000,0.000000,0.000000}%
\pgfsetfillcolor{currentfill}%
\pgfsetlinewidth{0.803000pt}%
\definecolor{currentstroke}{rgb}{0.000000,0.000000,0.000000}%
\pgfsetstrokecolor{currentstroke}%
\pgfsetdash{}{0pt}%
\pgfsys@defobject{currentmarker}{\pgfqpoint{0.000000in}{-0.048611in}}{\pgfqpoint{0.000000in}{0.000000in}}{%
\pgfpathmoveto{\pgfqpoint{0.000000in}{0.000000in}}%
\pgfpathlineto{\pgfqpoint{0.000000in}{-0.048611in}}%
\pgfusepath{stroke,fill}%
}%
\begin{pgfscope}%
\pgfsys@transformshift{1.783208in}{0.317817in}%
\pgfsys@useobject{currentmarker}{}%
\end{pgfscope}%
\end{pgfscope}%
\begin{pgfscope}%
\definecolor{textcolor}{rgb}{0.000000,0.000000,0.000000}%
\pgfsetstrokecolor{textcolor}%
\pgfsetfillcolor{textcolor}%
\pgftext[x=1.783208in,y=0.220595in,,top]{\color{textcolor}\rmfamily\fontsize{7.000000}{8.400000}\selectfont 15.2}%
\end{pgfscope}%
\begin{pgfscope}%
\pgfsetbuttcap%
\pgfsetroundjoin%
\definecolor{currentfill}{rgb}{0.000000,0.000000,0.000000}%
\pgfsetfillcolor{currentfill}%
\pgfsetlinewidth{0.803000pt}%
\definecolor{currentstroke}{rgb}{0.000000,0.000000,0.000000}%
\pgfsetstrokecolor{currentstroke}%
\pgfsetdash{}{0pt}%
\pgfsys@defobject{currentmarker}{\pgfqpoint{0.000000in}{-0.048611in}}{\pgfqpoint{0.000000in}{0.000000in}}{%
\pgfpathmoveto{\pgfqpoint{0.000000in}{0.000000in}}%
\pgfpathlineto{\pgfqpoint{0.000000in}{-0.048611in}}%
\pgfusepath{stroke,fill}%
}%
\begin{pgfscope}%
\pgfsys@transformshift{2.109025in}{0.317817in}%
\pgfsys@useobject{currentmarker}{}%
\end{pgfscope}%
\end{pgfscope}%
\begin{pgfscope}%
\definecolor{textcolor}{rgb}{0.000000,0.000000,0.000000}%
\pgfsetstrokecolor{textcolor}%
\pgfsetfillcolor{textcolor}%
\pgftext[x=2.109025in,y=0.220595in,,top]{\color{textcolor}\rmfamily\fontsize{7.000000}{8.400000}\selectfont 12.6}%
\end{pgfscope}%
\begin{pgfscope}%
\pgfsetbuttcap%
\pgfsetroundjoin%
\definecolor{currentfill}{rgb}{0.000000,0.000000,0.000000}%
\pgfsetfillcolor{currentfill}%
\pgfsetlinewidth{0.803000pt}%
\definecolor{currentstroke}{rgb}{0.000000,0.000000,0.000000}%
\pgfsetstrokecolor{currentstroke}%
\pgfsetdash{}{0pt}%
\pgfsys@defobject{currentmarker}{\pgfqpoint{0.000000in}{-0.048611in}}{\pgfqpoint{0.000000in}{0.000000in}}{%
\pgfpathmoveto{\pgfqpoint{0.000000in}{0.000000in}}%
\pgfpathlineto{\pgfqpoint{0.000000in}{-0.048611in}}%
\pgfusepath{stroke,fill}%
}%
\begin{pgfscope}%
\pgfsys@transformshift{2.434841in}{0.317817in}%
\pgfsys@useobject{currentmarker}{}%
\end{pgfscope}%
\end{pgfscope}%
\begin{pgfscope}%
\definecolor{textcolor}{rgb}{0.000000,0.000000,0.000000}%
\pgfsetstrokecolor{textcolor}%
\pgfsetfillcolor{textcolor}%
\pgftext[x=2.434841in,y=0.220595in,,top]{\color{textcolor}\rmfamily\fontsize{7.000000}{8.400000}\selectfont 10.8}%
\end{pgfscope}%
\begin{pgfscope}%
\pgfsetbuttcap%
\pgfsetroundjoin%
\definecolor{currentfill}{rgb}{0.000000,0.000000,0.000000}%
\pgfsetfillcolor{currentfill}%
\pgfsetlinewidth{0.803000pt}%
\definecolor{currentstroke}{rgb}{0.000000,0.000000,0.000000}%
\pgfsetstrokecolor{currentstroke}%
\pgfsetdash{}{0pt}%
\pgfsys@defobject{currentmarker}{\pgfqpoint{0.000000in}{-0.048611in}}{\pgfqpoint{0.000000in}{0.000000in}}{%
\pgfpathmoveto{\pgfqpoint{0.000000in}{0.000000in}}%
\pgfpathlineto{\pgfqpoint{0.000000in}{-0.048611in}}%
\pgfusepath{stroke,fill}%
}%
\begin{pgfscope}%
\pgfsys@transformshift{2.760658in}{0.317817in}%
\pgfsys@useobject{currentmarker}{}%
\end{pgfscope}%
\end{pgfscope}%
\begin{pgfscope}%
\definecolor{textcolor}{rgb}{0.000000,0.000000,0.000000}%
\pgfsetstrokecolor{textcolor}%
\pgfsetfillcolor{textcolor}%
\pgftext[x=2.760658in,y=0.220595in,,top]{\color{textcolor}\rmfamily\fontsize{7.000000}{8.400000}\selectfont 9.5}%
\end{pgfscope}%
\begin{pgfscope}%
\pgfsetbuttcap%
\pgfsetroundjoin%
\definecolor{currentfill}{rgb}{0.000000,0.000000,0.000000}%
\pgfsetfillcolor{currentfill}%
\pgfsetlinewidth{0.803000pt}%
\definecolor{currentstroke}{rgb}{0.000000,0.000000,0.000000}%
\pgfsetstrokecolor{currentstroke}%
\pgfsetdash{}{0pt}%
\pgfsys@defobject{currentmarker}{\pgfqpoint{0.000000in}{-0.048611in}}{\pgfqpoint{0.000000in}{0.000000in}}{%
\pgfpathmoveto{\pgfqpoint{0.000000in}{0.000000in}}%
\pgfpathlineto{\pgfqpoint{0.000000in}{-0.048611in}}%
\pgfusepath{stroke,fill}%
}%
\begin{pgfscope}%
\pgfsys@transformshift{3.086474in}{0.317817in}%
\pgfsys@useobject{currentmarker}{}%
\end{pgfscope}%
\end{pgfscope}%
\begin{pgfscope}%
\definecolor{textcolor}{rgb}{0.000000,0.000000,0.000000}%
\pgfsetstrokecolor{textcolor}%
\pgfsetfillcolor{textcolor}%
\pgftext[x=3.086474in,y=0.220595in,,top]{\color{textcolor}\rmfamily\fontsize{7.000000}{8.400000}\selectfont 8.4}%
\end{pgfscope}%
\begin{pgfscope}%
\definecolor{textcolor}{rgb}{0.000000,0.000000,0.000000}%
\pgfsetstrokecolor{textcolor}%
\pgfsetfillcolor{textcolor}%
\pgftext[x=1.783208in,y=0.078620in,,top]{\color{textcolor}\rmfamily\fontsize{7.000000}{8.400000}\selectfont Gate time [ms]}%
\end{pgfscope}%
\begin{pgfscope}%
\pgfsetrectcap%
\pgfsetmiterjoin%
\pgfsetlinewidth{0.803000pt}%
\definecolor{currentstroke}{rgb}{0.000000,0.000000,0.000000}%
\pgfsetstrokecolor{currentstroke}%
\pgfsetdash{}{0pt}%
\pgfpathmoveto{\pgfqpoint{3.086474in}{0.317817in}}%
\pgfpathlineto{\pgfqpoint{0.479942in}{0.317817in}}%
\pgfusepath{stroke}%
\end{pgfscope}%
\begin{pgfscope}%
\pgfpathrectangle{\pgfqpoint{0.479942in}{0.734383in}}{\pgfqpoint{2.606532in}{1.190189in}}%
\pgfusepath{clip}%
\pgfsetbuttcap%
\pgfsetroundjoin%
\pgfsetlinewidth{0.501875pt}%
\definecolor{currentstroke}{rgb}{0.933333,0.933333,0.933333}%
\pgfsetstrokecolor{currentstroke}%
\pgfsetdash{{1.850000pt}{0.800000pt}}{0.000000pt}%
\pgfpathmoveto{\pgfqpoint{0.479942in}{0.734383in}}%
\pgfpathlineto{\pgfqpoint{0.479942in}{1.924572in}}%
\pgfusepath{stroke}%
\end{pgfscope}%
\begin{pgfscope}%
\pgfsetbuttcap%
\pgfsetroundjoin%
\definecolor{currentfill}{rgb}{0.000000,0.000000,0.000000}%
\pgfsetfillcolor{currentfill}%
\pgfsetlinewidth{0.803000pt}%
\definecolor{currentstroke}{rgb}{0.000000,0.000000,0.000000}%
\pgfsetstrokecolor{currentstroke}%
\pgfsetdash{}{0pt}%
\pgfsys@defobject{currentmarker}{\pgfqpoint{0.000000in}{-0.048611in}}{\pgfqpoint{0.000000in}{0.000000in}}{%
\pgfpathmoveto{\pgfqpoint{0.000000in}{0.000000in}}%
\pgfpathlineto{\pgfqpoint{0.000000in}{-0.048611in}}%
\pgfusepath{stroke,fill}%
}%
\begin{pgfscope}%
\pgfsys@transformshift{0.479942in}{0.734383in}%
\pgfsys@useobject{currentmarker}{}%
\end{pgfscope}%
\end{pgfscope}%
\begin{pgfscope}%
\definecolor{textcolor}{rgb}{0.000000,0.000000,0.000000}%
\pgfsetstrokecolor{textcolor}%
\pgfsetfillcolor{textcolor}%
\pgftext[x=0.479942in,y=0.637161in,,top]{\color{textcolor}\rmfamily\fontsize{7.000000}{8.400000}\selectfont \(\displaystyle {0.2}\)}%
\end{pgfscope}%
\begin{pgfscope}%
\pgfpathrectangle{\pgfqpoint{0.479942in}{0.734383in}}{\pgfqpoint{2.606532in}{1.190189in}}%
\pgfusepath{clip}%
\pgfsetbuttcap%
\pgfsetroundjoin%
\pgfsetlinewidth{0.501875pt}%
\definecolor{currentstroke}{rgb}{0.933333,0.933333,0.933333}%
\pgfsetstrokecolor{currentstroke}%
\pgfsetdash{{1.850000pt}{0.800000pt}}{0.000000pt}%
\pgfpathmoveto{\pgfqpoint{0.805759in}{0.734383in}}%
\pgfpathlineto{\pgfqpoint{0.805759in}{1.924572in}}%
\pgfusepath{stroke}%
\end{pgfscope}%
\begin{pgfscope}%
\pgfsetbuttcap%
\pgfsetroundjoin%
\definecolor{currentfill}{rgb}{0.000000,0.000000,0.000000}%
\pgfsetfillcolor{currentfill}%
\pgfsetlinewidth{0.803000pt}%
\definecolor{currentstroke}{rgb}{0.000000,0.000000,0.000000}%
\pgfsetstrokecolor{currentstroke}%
\pgfsetdash{}{0pt}%
\pgfsys@defobject{currentmarker}{\pgfqpoint{0.000000in}{-0.048611in}}{\pgfqpoint{0.000000in}{0.000000in}}{%
\pgfpathmoveto{\pgfqpoint{0.000000in}{0.000000in}}%
\pgfpathlineto{\pgfqpoint{0.000000in}{-0.048611in}}%
\pgfusepath{stroke,fill}%
}%
\begin{pgfscope}%
\pgfsys@transformshift{0.805759in}{0.734383in}%
\pgfsys@useobject{currentmarker}{}%
\end{pgfscope}%
\end{pgfscope}%
\begin{pgfscope}%
\definecolor{textcolor}{rgb}{0.000000,0.000000,0.000000}%
\pgfsetstrokecolor{textcolor}%
\pgfsetfillcolor{textcolor}%
\pgftext[x=0.805759in,y=0.637161in,,top]{\color{textcolor}\rmfamily\fontsize{7.000000}{8.400000}\selectfont \(\displaystyle {0.4}\)}%
\end{pgfscope}%
\begin{pgfscope}%
\pgfpathrectangle{\pgfqpoint{0.479942in}{0.734383in}}{\pgfqpoint{2.606532in}{1.190189in}}%
\pgfusepath{clip}%
\pgfsetbuttcap%
\pgfsetroundjoin%
\pgfsetlinewidth{0.501875pt}%
\definecolor{currentstroke}{rgb}{0.933333,0.933333,0.933333}%
\pgfsetstrokecolor{currentstroke}%
\pgfsetdash{{1.850000pt}{0.800000pt}}{0.000000pt}%
\pgfpathmoveto{\pgfqpoint{1.131575in}{0.734383in}}%
\pgfpathlineto{\pgfqpoint{1.131575in}{1.924572in}}%
\pgfusepath{stroke}%
\end{pgfscope}%
\begin{pgfscope}%
\pgfsetbuttcap%
\pgfsetroundjoin%
\definecolor{currentfill}{rgb}{0.000000,0.000000,0.000000}%
\pgfsetfillcolor{currentfill}%
\pgfsetlinewidth{0.803000pt}%
\definecolor{currentstroke}{rgb}{0.000000,0.000000,0.000000}%
\pgfsetstrokecolor{currentstroke}%
\pgfsetdash{}{0pt}%
\pgfsys@defobject{currentmarker}{\pgfqpoint{0.000000in}{-0.048611in}}{\pgfqpoint{0.000000in}{0.000000in}}{%
\pgfpathmoveto{\pgfqpoint{0.000000in}{0.000000in}}%
\pgfpathlineto{\pgfqpoint{0.000000in}{-0.048611in}}%
\pgfusepath{stroke,fill}%
}%
\begin{pgfscope}%
\pgfsys@transformshift{1.131575in}{0.734383in}%
\pgfsys@useobject{currentmarker}{}%
\end{pgfscope}%
\end{pgfscope}%
\begin{pgfscope}%
\definecolor{textcolor}{rgb}{0.000000,0.000000,0.000000}%
\pgfsetstrokecolor{textcolor}%
\pgfsetfillcolor{textcolor}%
\pgftext[x=1.131575in,y=0.637161in,,top]{\color{textcolor}\rmfamily\fontsize{7.000000}{8.400000}\selectfont \(\displaystyle {0.6}\)}%
\end{pgfscope}%
\begin{pgfscope}%
\pgfpathrectangle{\pgfqpoint{0.479942in}{0.734383in}}{\pgfqpoint{2.606532in}{1.190189in}}%
\pgfusepath{clip}%
\pgfsetbuttcap%
\pgfsetroundjoin%
\pgfsetlinewidth{0.501875pt}%
\definecolor{currentstroke}{rgb}{0.933333,0.933333,0.933333}%
\pgfsetstrokecolor{currentstroke}%
\pgfsetdash{{1.850000pt}{0.800000pt}}{0.000000pt}%
\pgfpathmoveto{\pgfqpoint{1.457392in}{0.734383in}}%
\pgfpathlineto{\pgfqpoint{1.457392in}{1.924572in}}%
\pgfusepath{stroke}%
\end{pgfscope}%
\begin{pgfscope}%
\pgfsetbuttcap%
\pgfsetroundjoin%
\definecolor{currentfill}{rgb}{0.000000,0.000000,0.000000}%
\pgfsetfillcolor{currentfill}%
\pgfsetlinewidth{0.803000pt}%
\definecolor{currentstroke}{rgb}{0.000000,0.000000,0.000000}%
\pgfsetstrokecolor{currentstroke}%
\pgfsetdash{}{0pt}%
\pgfsys@defobject{currentmarker}{\pgfqpoint{0.000000in}{-0.048611in}}{\pgfqpoint{0.000000in}{0.000000in}}{%
\pgfpathmoveto{\pgfqpoint{0.000000in}{0.000000in}}%
\pgfpathlineto{\pgfqpoint{0.000000in}{-0.048611in}}%
\pgfusepath{stroke,fill}%
}%
\begin{pgfscope}%
\pgfsys@transformshift{1.457392in}{0.734383in}%
\pgfsys@useobject{currentmarker}{}%
\end{pgfscope}%
\end{pgfscope}%
\begin{pgfscope}%
\definecolor{textcolor}{rgb}{0.000000,0.000000,0.000000}%
\pgfsetstrokecolor{textcolor}%
\pgfsetfillcolor{textcolor}%
\pgftext[x=1.457392in,y=0.637161in,,top]{\color{textcolor}\rmfamily\fontsize{7.000000}{8.400000}\selectfont \(\displaystyle {0.8}\)}%
\end{pgfscope}%
\begin{pgfscope}%
\pgfpathrectangle{\pgfqpoint{0.479942in}{0.734383in}}{\pgfqpoint{2.606532in}{1.190189in}}%
\pgfusepath{clip}%
\pgfsetbuttcap%
\pgfsetroundjoin%
\pgfsetlinewidth{0.501875pt}%
\definecolor{currentstroke}{rgb}{0.933333,0.933333,0.933333}%
\pgfsetstrokecolor{currentstroke}%
\pgfsetdash{{1.850000pt}{0.800000pt}}{0.000000pt}%
\pgfpathmoveto{\pgfqpoint{1.783208in}{0.734383in}}%
\pgfpathlineto{\pgfqpoint{1.783208in}{1.924572in}}%
\pgfusepath{stroke}%
\end{pgfscope}%
\begin{pgfscope}%
\pgfsetbuttcap%
\pgfsetroundjoin%
\definecolor{currentfill}{rgb}{0.000000,0.000000,0.000000}%
\pgfsetfillcolor{currentfill}%
\pgfsetlinewidth{0.803000pt}%
\definecolor{currentstroke}{rgb}{0.000000,0.000000,0.000000}%
\pgfsetstrokecolor{currentstroke}%
\pgfsetdash{}{0pt}%
\pgfsys@defobject{currentmarker}{\pgfqpoint{0.000000in}{-0.048611in}}{\pgfqpoint{0.000000in}{0.000000in}}{%
\pgfpathmoveto{\pgfqpoint{0.000000in}{0.000000in}}%
\pgfpathlineto{\pgfqpoint{0.000000in}{-0.048611in}}%
\pgfusepath{stroke,fill}%
}%
\begin{pgfscope}%
\pgfsys@transformshift{1.783208in}{0.734383in}%
\pgfsys@useobject{currentmarker}{}%
\end{pgfscope}%
\end{pgfscope}%
\begin{pgfscope}%
\definecolor{textcolor}{rgb}{0.000000,0.000000,0.000000}%
\pgfsetstrokecolor{textcolor}%
\pgfsetfillcolor{textcolor}%
\pgftext[x=1.783208in,y=0.637161in,,top]{\color{textcolor}\rmfamily\fontsize{7.000000}{8.400000}\selectfont \(\displaystyle {1.0}\)}%
\end{pgfscope}%
\begin{pgfscope}%
\pgfpathrectangle{\pgfqpoint{0.479942in}{0.734383in}}{\pgfqpoint{2.606532in}{1.190189in}}%
\pgfusepath{clip}%
\pgfsetbuttcap%
\pgfsetroundjoin%
\pgfsetlinewidth{0.501875pt}%
\definecolor{currentstroke}{rgb}{0.933333,0.933333,0.933333}%
\pgfsetstrokecolor{currentstroke}%
\pgfsetdash{{1.850000pt}{0.800000pt}}{0.000000pt}%
\pgfpathmoveto{\pgfqpoint{2.109025in}{0.734383in}}%
\pgfpathlineto{\pgfqpoint{2.109025in}{1.924572in}}%
\pgfusepath{stroke}%
\end{pgfscope}%
\begin{pgfscope}%
\pgfsetbuttcap%
\pgfsetroundjoin%
\definecolor{currentfill}{rgb}{0.000000,0.000000,0.000000}%
\pgfsetfillcolor{currentfill}%
\pgfsetlinewidth{0.803000pt}%
\definecolor{currentstroke}{rgb}{0.000000,0.000000,0.000000}%
\pgfsetstrokecolor{currentstroke}%
\pgfsetdash{}{0pt}%
\pgfsys@defobject{currentmarker}{\pgfqpoint{0.000000in}{-0.048611in}}{\pgfqpoint{0.000000in}{0.000000in}}{%
\pgfpathmoveto{\pgfqpoint{0.000000in}{0.000000in}}%
\pgfpathlineto{\pgfqpoint{0.000000in}{-0.048611in}}%
\pgfusepath{stroke,fill}%
}%
\begin{pgfscope}%
\pgfsys@transformshift{2.109025in}{0.734383in}%
\pgfsys@useobject{currentmarker}{}%
\end{pgfscope}%
\end{pgfscope}%
\begin{pgfscope}%
\definecolor{textcolor}{rgb}{0.000000,0.000000,0.000000}%
\pgfsetstrokecolor{textcolor}%
\pgfsetfillcolor{textcolor}%
\pgftext[x=2.109025in,y=0.637161in,,top]{\color{textcolor}\rmfamily\fontsize{7.000000}{8.400000}\selectfont \(\displaystyle {1.2}\)}%
\end{pgfscope}%
\begin{pgfscope}%
\pgfpathrectangle{\pgfqpoint{0.479942in}{0.734383in}}{\pgfqpoint{2.606532in}{1.190189in}}%
\pgfusepath{clip}%
\pgfsetbuttcap%
\pgfsetroundjoin%
\pgfsetlinewidth{0.501875pt}%
\definecolor{currentstroke}{rgb}{0.933333,0.933333,0.933333}%
\pgfsetstrokecolor{currentstroke}%
\pgfsetdash{{1.850000pt}{0.800000pt}}{0.000000pt}%
\pgfpathmoveto{\pgfqpoint{2.434841in}{0.734383in}}%
\pgfpathlineto{\pgfqpoint{2.434841in}{1.924572in}}%
\pgfusepath{stroke}%
\end{pgfscope}%
\begin{pgfscope}%
\pgfsetbuttcap%
\pgfsetroundjoin%
\definecolor{currentfill}{rgb}{0.000000,0.000000,0.000000}%
\pgfsetfillcolor{currentfill}%
\pgfsetlinewidth{0.803000pt}%
\definecolor{currentstroke}{rgb}{0.000000,0.000000,0.000000}%
\pgfsetstrokecolor{currentstroke}%
\pgfsetdash{}{0pt}%
\pgfsys@defobject{currentmarker}{\pgfqpoint{0.000000in}{-0.048611in}}{\pgfqpoint{0.000000in}{0.000000in}}{%
\pgfpathmoveto{\pgfqpoint{0.000000in}{0.000000in}}%
\pgfpathlineto{\pgfqpoint{0.000000in}{-0.048611in}}%
\pgfusepath{stroke,fill}%
}%
\begin{pgfscope}%
\pgfsys@transformshift{2.434841in}{0.734383in}%
\pgfsys@useobject{currentmarker}{}%
\end{pgfscope}%
\end{pgfscope}%
\begin{pgfscope}%
\definecolor{textcolor}{rgb}{0.000000,0.000000,0.000000}%
\pgfsetstrokecolor{textcolor}%
\pgfsetfillcolor{textcolor}%
\pgftext[x=2.434841in,y=0.637161in,,top]{\color{textcolor}\rmfamily\fontsize{7.000000}{8.400000}\selectfont \(\displaystyle {1.4}\)}%
\end{pgfscope}%
\begin{pgfscope}%
\pgfpathrectangle{\pgfqpoint{0.479942in}{0.734383in}}{\pgfqpoint{2.606532in}{1.190189in}}%
\pgfusepath{clip}%
\pgfsetbuttcap%
\pgfsetroundjoin%
\pgfsetlinewidth{0.501875pt}%
\definecolor{currentstroke}{rgb}{0.933333,0.933333,0.933333}%
\pgfsetstrokecolor{currentstroke}%
\pgfsetdash{{1.850000pt}{0.800000pt}}{0.000000pt}%
\pgfpathmoveto{\pgfqpoint{2.760658in}{0.734383in}}%
\pgfpathlineto{\pgfqpoint{2.760658in}{1.924572in}}%
\pgfusepath{stroke}%
\end{pgfscope}%
\begin{pgfscope}%
\pgfsetbuttcap%
\pgfsetroundjoin%
\definecolor{currentfill}{rgb}{0.000000,0.000000,0.000000}%
\pgfsetfillcolor{currentfill}%
\pgfsetlinewidth{0.803000pt}%
\definecolor{currentstroke}{rgb}{0.000000,0.000000,0.000000}%
\pgfsetstrokecolor{currentstroke}%
\pgfsetdash{}{0pt}%
\pgfsys@defobject{currentmarker}{\pgfqpoint{0.000000in}{-0.048611in}}{\pgfqpoint{0.000000in}{0.000000in}}{%
\pgfpathmoveto{\pgfqpoint{0.000000in}{0.000000in}}%
\pgfpathlineto{\pgfqpoint{0.000000in}{-0.048611in}}%
\pgfusepath{stroke,fill}%
}%
\begin{pgfscope}%
\pgfsys@transformshift{2.760658in}{0.734383in}%
\pgfsys@useobject{currentmarker}{}%
\end{pgfscope}%
\end{pgfscope}%
\begin{pgfscope}%
\definecolor{textcolor}{rgb}{0.000000,0.000000,0.000000}%
\pgfsetstrokecolor{textcolor}%
\pgfsetfillcolor{textcolor}%
\pgftext[x=2.760658in,y=0.637161in,,top]{\color{textcolor}\rmfamily\fontsize{7.000000}{8.400000}\selectfont \(\displaystyle {1.6}\)}%
\end{pgfscope}%
\begin{pgfscope}%
\pgfpathrectangle{\pgfqpoint{0.479942in}{0.734383in}}{\pgfqpoint{2.606532in}{1.190189in}}%
\pgfusepath{clip}%
\pgfsetbuttcap%
\pgfsetroundjoin%
\pgfsetlinewidth{0.501875pt}%
\definecolor{currentstroke}{rgb}{0.933333,0.933333,0.933333}%
\pgfsetstrokecolor{currentstroke}%
\pgfsetdash{{1.850000pt}{0.800000pt}}{0.000000pt}%
\pgfpathmoveto{\pgfqpoint{3.086474in}{0.734383in}}%
\pgfpathlineto{\pgfqpoint{3.086474in}{1.924572in}}%
\pgfusepath{stroke}%
\end{pgfscope}%
\begin{pgfscope}%
\pgfsetbuttcap%
\pgfsetroundjoin%
\definecolor{currentfill}{rgb}{0.000000,0.000000,0.000000}%
\pgfsetfillcolor{currentfill}%
\pgfsetlinewidth{0.803000pt}%
\definecolor{currentstroke}{rgb}{0.000000,0.000000,0.000000}%
\pgfsetstrokecolor{currentstroke}%
\pgfsetdash{}{0pt}%
\pgfsys@defobject{currentmarker}{\pgfqpoint{0.000000in}{-0.048611in}}{\pgfqpoint{0.000000in}{0.000000in}}{%
\pgfpathmoveto{\pgfqpoint{0.000000in}{0.000000in}}%
\pgfpathlineto{\pgfqpoint{0.000000in}{-0.048611in}}%
\pgfusepath{stroke,fill}%
}%
\begin{pgfscope}%
\pgfsys@transformshift{3.086474in}{0.734383in}%
\pgfsys@useobject{currentmarker}{}%
\end{pgfscope}%
\end{pgfscope}%
\begin{pgfscope}%
\definecolor{textcolor}{rgb}{0.000000,0.000000,0.000000}%
\pgfsetstrokecolor{textcolor}%
\pgfsetfillcolor{textcolor}%
\pgftext[x=3.086474in,y=0.637161in,,top]{\color{textcolor}\rmfamily\fontsize{7.000000}{8.400000}\selectfont \(\displaystyle {1.8}\)}%
\end{pgfscope}%
\begin{pgfscope}%
\definecolor{textcolor}{rgb}{0.000000,0.000000,0.000000}%
\pgfsetstrokecolor{textcolor}%
\pgfsetfillcolor{textcolor}%
\pgftext[x=1.783208in,y=0.495186in,,top]{\color{textcolor}\rmfamily\fontsize{7.000000}{8.400000}\selectfont \(\displaystyle \Omega ~ / ~ J\)}%
\end{pgfscope}%
\begin{pgfscope}%
\pgfpathrectangle{\pgfqpoint{0.479942in}{0.734383in}}{\pgfqpoint{2.606532in}{1.190189in}}%
\pgfusepath{clip}%
\pgfsetbuttcap%
\pgfsetroundjoin%
\pgfsetlinewidth{0.501875pt}%
\definecolor{currentstroke}{rgb}{0.933333,0.933333,0.933333}%
\pgfsetstrokecolor{currentstroke}%
\pgfsetdash{{1.850000pt}{0.800000pt}}{0.000000pt}%
\pgfpathmoveto{\pgfqpoint{0.479942in}{0.734383in}}%
\pgfpathlineto{\pgfqpoint{3.086474in}{0.734383in}}%
\pgfusepath{stroke}%
\end{pgfscope}%
\begin{pgfscope}%
\pgfsetbuttcap%
\pgfsetroundjoin%
\definecolor{currentfill}{rgb}{0.000000,0.000000,0.000000}%
\pgfsetfillcolor{currentfill}%
\pgfsetlinewidth{0.803000pt}%
\definecolor{currentstroke}{rgb}{0.000000,0.000000,0.000000}%
\pgfsetstrokecolor{currentstroke}%
\pgfsetdash{}{0pt}%
\pgfsys@defobject{currentmarker}{\pgfqpoint{-0.048611in}{0.000000in}}{\pgfqpoint{-0.000000in}{0.000000in}}{%
\pgfpathmoveto{\pgfqpoint{-0.000000in}{0.000000in}}%
\pgfpathlineto{\pgfqpoint{-0.048611in}{0.000000in}}%
\pgfusepath{stroke,fill}%
}%
\begin{pgfscope}%
\pgfsys@transformshift{0.479942in}{0.734383in}%
\pgfsys@useobject{currentmarker}{}%
\end{pgfscope}%
\end{pgfscope}%
\begin{pgfscope}%
\definecolor{textcolor}{rgb}{0.000000,0.000000,0.000000}%
\pgfsetstrokecolor{textcolor}%
\pgfsetfillcolor{textcolor}%
\pgftext[x=0.183645in, y=0.700626in, left, base]{\color{textcolor}\rmfamily\fontsize{7.000000}{8.400000}\selectfont \(\displaystyle {0.75}\)}%
\end{pgfscope}%
\begin{pgfscope}%
\pgfpathrectangle{\pgfqpoint{0.479942in}{0.734383in}}{\pgfqpoint{2.606532in}{1.190189in}}%
\pgfusepath{clip}%
\pgfsetbuttcap%
\pgfsetroundjoin%
\pgfsetlinewidth{0.501875pt}%
\definecolor{currentstroke}{rgb}{0.933333,0.933333,0.933333}%
\pgfsetstrokecolor{currentstroke}%
\pgfsetdash{{1.850000pt}{0.800000pt}}{0.000000pt}%
\pgfpathmoveto{\pgfqpoint{0.479942in}{0.972421in}}%
\pgfpathlineto{\pgfqpoint{3.086474in}{0.972421in}}%
\pgfusepath{stroke}%
\end{pgfscope}%
\begin{pgfscope}%
\pgfsetbuttcap%
\pgfsetroundjoin%
\definecolor{currentfill}{rgb}{0.000000,0.000000,0.000000}%
\pgfsetfillcolor{currentfill}%
\pgfsetlinewidth{0.803000pt}%
\definecolor{currentstroke}{rgb}{0.000000,0.000000,0.000000}%
\pgfsetstrokecolor{currentstroke}%
\pgfsetdash{}{0pt}%
\pgfsys@defobject{currentmarker}{\pgfqpoint{-0.048611in}{0.000000in}}{\pgfqpoint{-0.000000in}{0.000000in}}{%
\pgfpathmoveto{\pgfqpoint{-0.000000in}{0.000000in}}%
\pgfpathlineto{\pgfqpoint{-0.048611in}{0.000000in}}%
\pgfusepath{stroke,fill}%
}%
\begin{pgfscope}%
\pgfsys@transformshift{0.479942in}{0.972421in}%
\pgfsys@useobject{currentmarker}{}%
\end{pgfscope}%
\end{pgfscope}%
\begin{pgfscope}%
\definecolor{textcolor}{rgb}{0.000000,0.000000,0.000000}%
\pgfsetstrokecolor{textcolor}%
\pgfsetfillcolor{textcolor}%
\pgftext[x=0.183645in, y=0.938664in, left, base]{\color{textcolor}\rmfamily\fontsize{7.000000}{8.400000}\selectfont \(\displaystyle {0.80}\)}%
\end{pgfscope}%
\begin{pgfscope}%
\pgfpathrectangle{\pgfqpoint{0.479942in}{0.734383in}}{\pgfqpoint{2.606532in}{1.190189in}}%
\pgfusepath{clip}%
\pgfsetbuttcap%
\pgfsetroundjoin%
\pgfsetlinewidth{0.501875pt}%
\definecolor{currentstroke}{rgb}{0.933333,0.933333,0.933333}%
\pgfsetstrokecolor{currentstroke}%
\pgfsetdash{{1.850000pt}{0.800000pt}}{0.000000pt}%
\pgfpathmoveto{\pgfqpoint{0.479942in}{1.210459in}}%
\pgfpathlineto{\pgfqpoint{3.086474in}{1.210459in}}%
\pgfusepath{stroke}%
\end{pgfscope}%
\begin{pgfscope}%
\pgfsetbuttcap%
\pgfsetroundjoin%
\definecolor{currentfill}{rgb}{0.000000,0.000000,0.000000}%
\pgfsetfillcolor{currentfill}%
\pgfsetlinewidth{0.803000pt}%
\definecolor{currentstroke}{rgb}{0.000000,0.000000,0.000000}%
\pgfsetstrokecolor{currentstroke}%
\pgfsetdash{}{0pt}%
\pgfsys@defobject{currentmarker}{\pgfqpoint{-0.048611in}{0.000000in}}{\pgfqpoint{-0.000000in}{0.000000in}}{%
\pgfpathmoveto{\pgfqpoint{-0.000000in}{0.000000in}}%
\pgfpathlineto{\pgfqpoint{-0.048611in}{0.000000in}}%
\pgfusepath{stroke,fill}%
}%
\begin{pgfscope}%
\pgfsys@transformshift{0.479942in}{1.210459in}%
\pgfsys@useobject{currentmarker}{}%
\end{pgfscope}%
\end{pgfscope}%
\begin{pgfscope}%
\definecolor{textcolor}{rgb}{0.000000,0.000000,0.000000}%
\pgfsetstrokecolor{textcolor}%
\pgfsetfillcolor{textcolor}%
\pgftext[x=0.183645in, y=1.176701in, left, base]{\color{textcolor}\rmfamily\fontsize{7.000000}{8.400000}\selectfont \(\displaystyle {0.85}\)}%
\end{pgfscope}%
\begin{pgfscope}%
\pgfpathrectangle{\pgfqpoint{0.479942in}{0.734383in}}{\pgfqpoint{2.606532in}{1.190189in}}%
\pgfusepath{clip}%
\pgfsetbuttcap%
\pgfsetroundjoin%
\pgfsetlinewidth{0.501875pt}%
\definecolor{currentstroke}{rgb}{0.933333,0.933333,0.933333}%
\pgfsetstrokecolor{currentstroke}%
\pgfsetdash{{1.850000pt}{0.800000pt}}{0.000000pt}%
\pgfpathmoveto{\pgfqpoint{0.479942in}{1.448497in}}%
\pgfpathlineto{\pgfqpoint{3.086474in}{1.448497in}}%
\pgfusepath{stroke}%
\end{pgfscope}%
\begin{pgfscope}%
\pgfsetbuttcap%
\pgfsetroundjoin%
\definecolor{currentfill}{rgb}{0.000000,0.000000,0.000000}%
\pgfsetfillcolor{currentfill}%
\pgfsetlinewidth{0.803000pt}%
\definecolor{currentstroke}{rgb}{0.000000,0.000000,0.000000}%
\pgfsetstrokecolor{currentstroke}%
\pgfsetdash{}{0pt}%
\pgfsys@defobject{currentmarker}{\pgfqpoint{-0.048611in}{0.000000in}}{\pgfqpoint{-0.000000in}{0.000000in}}{%
\pgfpathmoveto{\pgfqpoint{-0.000000in}{0.000000in}}%
\pgfpathlineto{\pgfqpoint{-0.048611in}{0.000000in}}%
\pgfusepath{stroke,fill}%
}%
\begin{pgfscope}%
\pgfsys@transformshift{0.479942in}{1.448497in}%
\pgfsys@useobject{currentmarker}{}%
\end{pgfscope}%
\end{pgfscope}%
\begin{pgfscope}%
\definecolor{textcolor}{rgb}{0.000000,0.000000,0.000000}%
\pgfsetstrokecolor{textcolor}%
\pgfsetfillcolor{textcolor}%
\pgftext[x=0.183645in, y=1.414739in, left, base]{\color{textcolor}\rmfamily\fontsize{7.000000}{8.400000}\selectfont \(\displaystyle {0.90}\)}%
\end{pgfscope}%
\begin{pgfscope}%
\pgfpathrectangle{\pgfqpoint{0.479942in}{0.734383in}}{\pgfqpoint{2.606532in}{1.190189in}}%
\pgfusepath{clip}%
\pgfsetbuttcap%
\pgfsetroundjoin%
\pgfsetlinewidth{0.501875pt}%
\definecolor{currentstroke}{rgb}{0.933333,0.933333,0.933333}%
\pgfsetstrokecolor{currentstroke}%
\pgfsetdash{{1.850000pt}{0.800000pt}}{0.000000pt}%
\pgfpathmoveto{\pgfqpoint{0.479942in}{1.686535in}}%
\pgfpathlineto{\pgfqpoint{3.086474in}{1.686535in}}%
\pgfusepath{stroke}%
\end{pgfscope}%
\begin{pgfscope}%
\pgfsetbuttcap%
\pgfsetroundjoin%
\definecolor{currentfill}{rgb}{0.000000,0.000000,0.000000}%
\pgfsetfillcolor{currentfill}%
\pgfsetlinewidth{0.803000pt}%
\definecolor{currentstroke}{rgb}{0.000000,0.000000,0.000000}%
\pgfsetstrokecolor{currentstroke}%
\pgfsetdash{}{0pt}%
\pgfsys@defobject{currentmarker}{\pgfqpoint{-0.048611in}{0.000000in}}{\pgfqpoint{-0.000000in}{0.000000in}}{%
\pgfpathmoveto{\pgfqpoint{-0.000000in}{0.000000in}}%
\pgfpathlineto{\pgfqpoint{-0.048611in}{0.000000in}}%
\pgfusepath{stroke,fill}%
}%
\begin{pgfscope}%
\pgfsys@transformshift{0.479942in}{1.686535in}%
\pgfsys@useobject{currentmarker}{}%
\end{pgfscope}%
\end{pgfscope}%
\begin{pgfscope}%
\definecolor{textcolor}{rgb}{0.000000,0.000000,0.000000}%
\pgfsetstrokecolor{textcolor}%
\pgfsetfillcolor{textcolor}%
\pgftext[x=0.183645in, y=1.652777in, left, base]{\color{textcolor}\rmfamily\fontsize{7.000000}{8.400000}\selectfont \(\displaystyle {0.95}\)}%
\end{pgfscope}%
\begin{pgfscope}%
\pgfpathrectangle{\pgfqpoint{0.479942in}{0.734383in}}{\pgfqpoint{2.606532in}{1.190189in}}%
\pgfusepath{clip}%
\pgfsetbuttcap%
\pgfsetroundjoin%
\pgfsetlinewidth{0.501875pt}%
\definecolor{currentstroke}{rgb}{0.933333,0.933333,0.933333}%
\pgfsetstrokecolor{currentstroke}%
\pgfsetdash{{1.850000pt}{0.800000pt}}{0.000000pt}%
\pgfpathmoveto{\pgfqpoint{0.479942in}{1.924572in}}%
\pgfpathlineto{\pgfqpoint{3.086474in}{1.924572in}}%
\pgfusepath{stroke}%
\end{pgfscope}%
\begin{pgfscope}%
\pgfsetbuttcap%
\pgfsetroundjoin%
\definecolor{currentfill}{rgb}{0.000000,0.000000,0.000000}%
\pgfsetfillcolor{currentfill}%
\pgfsetlinewidth{0.803000pt}%
\definecolor{currentstroke}{rgb}{0.000000,0.000000,0.000000}%
\pgfsetstrokecolor{currentstroke}%
\pgfsetdash{}{0pt}%
\pgfsys@defobject{currentmarker}{\pgfqpoint{-0.048611in}{0.000000in}}{\pgfqpoint{-0.000000in}{0.000000in}}{%
\pgfpathmoveto{\pgfqpoint{-0.000000in}{0.000000in}}%
\pgfpathlineto{\pgfqpoint{-0.048611in}{0.000000in}}%
\pgfusepath{stroke,fill}%
}%
\begin{pgfscope}%
\pgfsys@transformshift{0.479942in}{1.924572in}%
\pgfsys@useobject{currentmarker}{}%
\end{pgfscope}%
\end{pgfscope}%
\begin{pgfscope}%
\definecolor{textcolor}{rgb}{0.000000,0.000000,0.000000}%
\pgfsetstrokecolor{textcolor}%
\pgfsetfillcolor{textcolor}%
\pgftext[x=0.183645in, y=1.890815in, left, base]{\color{textcolor}\rmfamily\fontsize{7.000000}{8.400000}\selectfont \(\displaystyle {1.00}\)}%
\end{pgfscope}%
\begin{pgfscope}%
\definecolor{textcolor}{rgb}{0.000000,0.000000,0.000000}%
\pgfsetstrokecolor{textcolor}%
\pgfsetfillcolor{textcolor}%
\pgftext[x=0.128089in,y=1.329478in,,bottom,rotate=90.000000]{\color{textcolor}\rmfamily\fontsize{7.000000}{8.400000}\selectfont Toffoli gate fidelity}%
\end{pgfscope}%
\begin{pgfscope}%
\pgfpathrectangle{\pgfqpoint{0.479942in}{0.734383in}}{\pgfqpoint{2.606532in}{1.190189in}}%
\pgfusepath{clip}%
\pgfsetrectcap%
\pgfsetroundjoin%
\pgfsetlinewidth{1.003750pt}%
\definecolor{currentstroke}{rgb}{0.000000,0.000000,0.000000}%
\pgfsetstrokecolor{currentstroke}%
\pgfsetdash{}{0pt}%
\pgfpathmoveto{\pgfqpoint{0.478275in}{1.903581in}}%
\pgfpathlineto{\pgfqpoint{0.483014in}{1.911083in}}%
\pgfpathlineto{\pgfqpoint{0.499451in}{1.878655in}}%
\pgfpathlineto{\pgfqpoint{0.516288in}{1.855681in}}%
\pgfpathlineto{\pgfqpoint{0.533526in}{1.861645in}}%
\pgfpathlineto{\pgfqpoint{0.551166in}{1.897279in}}%
\pgfpathlineto{\pgfqpoint{0.569205in}{1.913294in}}%
\pgfpathlineto{\pgfqpoint{0.587646in}{1.876581in}}%
\pgfpathlineto{\pgfqpoint{0.606487in}{1.838960in}}%
\pgfpathlineto{\pgfqpoint{0.625729in}{1.819626in}}%
\pgfpathlineto{\pgfqpoint{0.645372in}{1.823239in}}%
\pgfpathlineto{\pgfqpoint{0.665416in}{1.857173in}}%
\pgfpathlineto{\pgfqpoint{0.685860in}{1.903436in}}%
\pgfpathlineto{\pgfqpoint{0.706705in}{1.918316in}}%
\pgfpathlineto{\pgfqpoint{0.727951in}{1.891285in}}%
\pgfpathlineto{\pgfqpoint{0.749598in}{1.840501in}}%
\pgfpathlineto{\pgfqpoint{0.771645in}{1.793451in}}%
\pgfpathlineto{\pgfqpoint{0.794094in}{1.758747in}}%
\pgfpathlineto{\pgfqpoint{0.816943in}{1.733295in}}%
\pgfpathlineto{\pgfqpoint{0.840193in}{1.721708in}}%
\pgfpathlineto{\pgfqpoint{0.863843in}{1.729117in}}%
\pgfpathlineto{\pgfqpoint{0.887895in}{1.759748in}}%
\pgfpathlineto{\pgfqpoint{0.912347in}{1.809637in}}%
\pgfpathlineto{\pgfqpoint{0.937200in}{1.863672in}}%
\pgfpathlineto{\pgfqpoint{0.962453in}{1.904875in}}%
\pgfpathlineto{\pgfqpoint{0.988108in}{1.920336in}}%
\pgfpathlineto{\pgfqpoint{1.014163in}{1.907365in}}%
\pgfpathlineto{\pgfqpoint{1.040619in}{1.869494in}}%
\pgfpathlineto{\pgfqpoint{1.067475in}{1.815483in}}%
\pgfpathlineto{\pgfqpoint{1.094733in}{1.755472in}}%
\pgfpathlineto{\pgfqpoint{1.122391in}{1.694121in}}%
\pgfpathlineto{\pgfqpoint{1.150450in}{1.637477in}}%
\pgfpathlineto{\pgfqpoint{1.178910in}{1.586360in}}%
\pgfpathlineto{\pgfqpoint{1.207771in}{1.537512in}}%
\pgfpathlineto{\pgfqpoint{1.237032in}{1.492987in}}%
\pgfpathlineto{\pgfqpoint{1.266694in}{1.450191in}}%
\pgfpathlineto{\pgfqpoint{1.296757in}{1.410285in}}%
\pgfpathlineto{\pgfqpoint{1.327221in}{1.376369in}}%
\pgfpathlineto{\pgfqpoint{1.358085in}{1.351518in}}%
\pgfpathlineto{\pgfqpoint{1.389351in}{1.333571in}}%
\pgfpathlineto{\pgfqpoint{1.421016in}{1.328265in}}%
\pgfpathlineto{\pgfqpoint{1.453083in}{1.336922in}}%
\pgfpathlineto{\pgfqpoint{1.485551in}{1.355455in}}%
\pgfpathlineto{\pgfqpoint{1.518419in}{1.388713in}}%
\pgfpathlineto{\pgfqpoint{1.551688in}{1.430070in}}%
\pgfpathlineto{\pgfqpoint{1.585358in}{1.481366in}}%
\pgfpathlineto{\pgfqpoint{1.619429in}{1.538493in}}%
\pgfpathlineto{\pgfqpoint{1.653900in}{1.596452in}}%
\pgfpathlineto{\pgfqpoint{1.688772in}{1.656785in}}%
\pgfpathlineto{\pgfqpoint{1.724045in}{1.712898in}}%
\pgfpathlineto{\pgfqpoint{1.759719in}{1.763406in}}%
\pgfpathlineto{\pgfqpoint{1.795793in}{1.806659in}}%
\pgfpathlineto{\pgfqpoint{1.832268in}{1.842089in}}%
\pgfpathlineto{\pgfqpoint{1.869145in}{1.867464in}}%
\pgfpathlineto{\pgfqpoint{1.906421in}{1.882818in}}%
\pgfpathlineto{\pgfqpoint{1.944099in}{1.888339in}}%
\pgfpathlineto{\pgfqpoint{1.982177in}{1.884430in}}%
\pgfpathlineto{\pgfqpoint{2.020656in}{1.870543in}}%
\pgfpathlineto{\pgfqpoint{2.059536in}{1.848801in}}%
\pgfpathlineto{\pgfqpoint{2.098817in}{1.820543in}}%
\pgfpathlineto{\pgfqpoint{2.138498in}{1.785092in}}%
\pgfpathlineto{\pgfqpoint{2.178580in}{1.744528in}}%
\pgfpathlineto{\pgfqpoint{2.219063in}{1.701175in}}%
\pgfpathlineto{\pgfqpoint{2.259947in}{1.652117in}}%
\pgfpathlineto{\pgfqpoint{2.301232in}{1.603638in}}%
\pgfpathlineto{\pgfqpoint{2.342917in}{1.551005in}}%
\pgfpathlineto{\pgfqpoint{2.385003in}{1.499427in}}%
\pgfpathlineto{\pgfqpoint{2.427490in}{1.448149in}}%
\pgfpathlineto{\pgfqpoint{2.470377in}{1.394530in}}%
\pgfpathlineto{\pgfqpoint{2.513666in}{1.343114in}}%
\pgfpathlineto{\pgfqpoint{2.557355in}{1.293161in}}%
\pgfpathlineto{\pgfqpoint{2.601445in}{1.243816in}}%
\pgfpathlineto{\pgfqpoint{2.645935in}{1.193901in}}%
\pgfpathlineto{\pgfqpoint{2.690827in}{1.142664in}}%
\pgfpathlineto{\pgfqpoint{2.736119in}{1.096504in}}%
\pgfpathlineto{\pgfqpoint{2.781812in}{1.049218in}}%
\pgfpathlineto{\pgfqpoint{2.827906in}{1.003772in}}%
\pgfpathlineto{\pgfqpoint{2.874400in}{0.959011in}}%
\pgfpathlineto{\pgfqpoint{2.921295in}{0.914508in}}%
\pgfpathlineto{\pgfqpoint{2.968591in}{0.869450in}}%
\pgfpathlineto{\pgfqpoint{3.016288in}{0.824834in}}%
\pgfpathlineto{\pgfqpoint{3.064386in}{0.782870in}}%
\pgfpathlineto{\pgfqpoint{3.088141in}{0.762684in}}%
\pgfusepath{stroke}%
\end{pgfscope}%
\begin{pgfscope}%
\pgfsetrectcap%
\pgfsetmiterjoin%
\pgfsetlinewidth{0.803000pt}%
\definecolor{currentstroke}{rgb}{0.000000,0.000000,0.000000}%
\pgfsetstrokecolor{currentstroke}%
\pgfsetdash{}{0pt}%
\pgfpathmoveto{\pgfqpoint{0.479942in}{0.734383in}}%
\pgfpathlineto{\pgfqpoint{0.479942in}{1.924572in}}%
\pgfusepath{stroke}%
\end{pgfscope}%
\begin{pgfscope}%
\pgfsetrectcap%
\pgfsetmiterjoin%
\pgfsetlinewidth{0.803000pt}%
\definecolor{currentstroke}{rgb}{0.000000,0.000000,0.000000}%
\pgfsetstrokecolor{currentstroke}%
\pgfsetdash{}{0pt}%
\pgfpathmoveto{\pgfqpoint{3.086474in}{0.734383in}}%
\pgfpathlineto{\pgfqpoint{3.086474in}{1.924572in}}%
\pgfusepath{stroke}%
\end{pgfscope}%
\begin{pgfscope}%
\pgfsetrectcap%
\pgfsetmiterjoin%
\pgfsetlinewidth{0.803000pt}%
\definecolor{currentstroke}{rgb}{0.000000,0.000000,0.000000}%
\pgfsetstrokecolor{currentstroke}%
\pgfsetdash{}{0pt}%
\pgfpathmoveto{\pgfqpoint{0.479942in}{0.734383in}}%
\pgfpathlineto{\pgfqpoint{3.086474in}{0.734383in}}%
\pgfusepath{stroke}%
\end{pgfscope}%
\begin{pgfscope}%
\pgfsetrectcap%
\pgfsetmiterjoin%
\pgfsetlinewidth{0.803000pt}%
\definecolor{currentstroke}{rgb}{0.000000,0.000000,0.000000}%
\pgfsetstrokecolor{currentstroke}%
\pgfsetdash{}{0pt}%
\pgfpathmoveto{\pgfqpoint{0.479942in}{1.924572in}}%
\pgfpathlineto{\pgfqpoint{3.086474in}{1.924572in}}%
\pgfusepath{stroke}%
\end{pgfscope}%
\end{pgfpicture}%
\makeatother%
\endgroup%